\documentclass[12pt]{article}
\usepackage[margin=0.8in]{geometry}
\usepackage[utf8]{inputenc}
\usepackage{graphicx}
\usepackage{amssymb}
\usepackage{amsmath}
\usepackage{amsfonts}
\usepackage{xcolor}
\usepackage{url}
\usepackage{float}
\usepackage{hyperref}
\usepackage{soul}
\hypersetup{
  colorlinks   = true, 
  urlcolor     = blue, 
  linkcolor    = black, 
  citecolor   = black 
}

\usepackage[numbers,sort&compress]{natbib}
\let\OLDthebibliography\thebibliography
\renewcommand\thebibliography[1]{
  \OLDthebibliography{#1}
  \setlength{\parskip}{4pt}
  \setlength{\itemsep}{0pt plus 0.3ex}
}

\newcommand{\tautau}{{\tau^+\tau^-}}
\newcommand{\gamgam}{{\gamma\gamma}}
\newcommand{\bbbar}{{b \bar b}}
\newcommand{\gev}{\,\mathrm{GeV}}
\newcommand{\tev}{\,\mathrm{TeV}}
\newcommand{\tb}{{\tan\beta}}
\newcommand\mh[1]{m_{h_{#1}}}
\newcommand\SW{s_\mathrm{w}}
\newcommand\CW{c_\mathrm{w}}
\newcommand\MW{M_W}
\newcommand\MZ{M_Z}
\newcommand\al{\alpha}
\newcommand\be{\beta}
\newcommand\ga{\gamma}
\newcommand\MHp{m_{H^\pm}}
\newcommand{\Hpm}{H^\pm}
\newcommand{\he}{\ensuremath{h_{95}}}
\newcommand{\hz}{\ensuremath{h_{125}}}
\newcommand{\br}{\mathrm{BR}}

\newcommand\refeq[1]{Eq.~(\ref{#1})}

\newcommand\refta[1]{Tab.~\ref{#1}}
\newcommand\refse[1]{Sect.~\ref{#1}}

\newcommand\citere[1]{Ref.~\cite{#1}}
\newcommand\citeres[1]{Refs.~\cite{#1}}
\def\reffi#1{\mbox{Fig.~\ref{#1}}}

\newcommand{\hto}[1]{{\color{black}#1}}

\newcommand{\GWn}[1]{{\color{black}#1}}

\newcommand{\GW}[1]{{\color{black}#1}}
\newcommand{\tho}[1]{{\color{black}#1}}

\begin{document}

\thispagestyle{empty}

\def\thefootnote{\fnsymbol{footnote}}

\begin{flushright}
  DESY 22-057 ~~\\ IFT-UAM/CSIC--22--033 ~~
\end{flushright}

\vspace*{1.4cm}

\begin{center}

{\Large \textbf{Mounting evidence
for a 95 GeV Higgs boson}}

\vspace{1cm}

{\sc
T.~Biekötter\footnote{thomas.biekoetter@desy.de}$^1$,
S.~Heinemeyer\footnote{Sven.Heinemeyer@cern.ch}$^{2}$
and  G.~Weiglein\footnote{georg.weiglein@desy.de}$^{1,3}$
}

\vspace*{.7cm}

{\sl
$^1$Deutsches Elektronen-Synchrotron DESY,
Notkestr.~85,  22607  Hamburg,  Germany

\vspace{0.1cm}

$^2$Instituto de F\'isica Te\'orica UAM-CSIC, Cantoblanco, 28049,
      Madrid, Spain

\vspace{0.1cm}

$^3$II.\  Institut f\"ur  Theoretische  Physik, Universit\"at  Hamburg,
Luruper Chaussee 149, 22761 Hamburg, Germany
}

\end{center}

\vspace*{0.1cm}

\begin{abstract}
In 2018 CMS reported an excess
in the light Higgs-boson search
in the diphoton decay mode at
about $95 \gev$
based on Run~1 and first year Run~2 data.
The combined local significance of the
excess was $2.8\,\sigma$.
The excess is compatible with the
limits obtained in the ATLAS searches
\GW{from} the diphoton search channel.
Recently, CMS reported another local
excess with a significance of
$3.1\,\sigma$ in the light Higgs-boson search
in the di-tau final state, which is
compatible with the interpretation
of a Higgs boson with a mass of
about $95\gev$.
We \GW{show that the observed results can be interpreted as manifestations of} 
a Higgs boson in the Two-Higgs Doublet
Model with an additional real
singlet (N2HDM).
We find that the lightest Higgs boson
of the N2HDM can fit
both excesses simultaneously,
while the second-lightest state is
such that it satisfies the Higgs-boson measurements at
$125 \gev$, and the full Higgs-boson
sector is compatible with all Higgs exclusion
bounds from the searches at
LEP, the Tevatron and the LHC as
well as with other theoretical and
experimental constraints.
Finally, we demonstrate that it
is furthermore possible to
accommodate the excesses observed
by CMS \GW{in the two search channels} together
with a local $2.3\,\sigma$ excess
in the $b \bar b$ final state
observed at LEP in the same mass range.
\end{abstract}

\def\thefootnote{\arabic{footnote}}
\setcounter{page}{0}
\setcounter{footnote}{0}

\newpage

\section{Introduction}
In the year 2012 the ATLAS and CMS collaborations 
discovered a new
particle\GW{~\cite{Aad:2012tfa,Chatrchyan:2012xdj,Khachatryan:2016vau}}
that -- within the present
theoretical and experimental uncertainties -- is
consistent with the 
\GW{predictions for the Higgs boson of the}
Standard Model~(SM) 
at a mass
of \GW{about} 
$125 \gev$, \GW{but is also compatible with the predictions of a wide variety of extensions of the SM.
While no} conclusive signs of physics beyond the~SM
\tho{(BSM)} have been found so far at the LHC,
\GW{both the measurements of the properties of the discovered state at $125\gev$ (its couplings are}
known up to now
to an experimental precision of roughly $20\%$) 
\GW{and the existing limits from the searches for new particles}
leave significant room for interpretations in
models of physics beyond the SM.
Many BSM models feature
extended Higgs-boson sectors. Consequently, one of the main tasks of the
LHC Run~3 and beyond will be to determine whether
the observed scalar boson forms part of the Higgs sector of an extended
model.
Extended Higgs-boson sectors naturally contain additional Higgs bosons with
masses larger than $125 \gev$. However, many extensions also offer the
possibilty of additional Higgs bosons that are {\em lighter} than
$125 \gev$.
Accordingly, the search for
light additional Higgs bosons is of crucial importance for exploring the
underlying physics of electroweak symmetry breaking.

Searches for Higgs bosons below $125 \gev$ have been performed at
LEP~\cite{Abbiendi:2002qp,Barate:2003sz,Schael:2006cr},
the Tevatron~\cite{Group:2012zca} and
the LHC~\cite{CMS:2017yta,Sirunyan:2018aui,
Sirunyan:2018zut,ATLAS:2018xad,CMS-PAS-HIG-21-001}.
Results based on the first year of CMS~Run\,2 data for Higgs-boson
searches in the diphoton
final state show a local excess of about~$3\,\sigma$ at a mass of
$95\gev$~\cite{Sirunyan:2018aui},
which received considerable attention
also in view of the fact that a similar excess
of~$2\,\sigma$ occurred in the Run\,1 data at a comparable
mass~\cite{CMS:2015ocq}.
Combining 7, 8~and first year $13 \tev$ data (and assuming that the $gg$ production dominates)
the excess is most pronounced at a mass
of $95.3$ 
with a local significance of
$2.8\,\sigma$.
From the excess of events
CMS obtained a signal strength of
\begin{equation}
\mu_{\gamgam}^{\rm exp} =\frac{\sigma^{\rm exp} \left( gg \to \phi \to \gamma\gamma \right)}
         {\sigma^{\rm SM}\left( gg \to H \to \gamma\gamma \right)}
     = 0.6 \pm 0.2 \ .
\label{muCMS}
\end{equation}
Here the SM prediction, \tho{$\sigma^{\rm SM}$},
is evaluated for a SM Higgs-boson mass of
$95.3 \gev$.
First Run\,2~results from~ATLAS
with~$80$\,fb$^{-1}$ in the~$\gamgam$~searches below~$125$\,GeV were
reported in 2018~\cite{ATLAS:2018xad}. No significant
excess above the~SM~expectation was observed in the mass range between
$65$~and $110 \gev$.
However, the limit on cross section times branching ratio obtained in the
diphoton final state by ATLAS is not only well above $\mu^{\rm exp}_\gamgam$,
but even weaker than the corresponding upper limit obtained by CMS at and around
$95 \gev$. This was illustrated in Fig.~1
of~\citere{Heinemeyer:2018wzl} (based on the Run~1 and first year Run~2
data).

\GW{Searches for a low-mass Higgs boson that were previously carried out at 
LEP resulted in}
a $2.3\,\sigma$ local excess
observed in the~$e^+e^-\to Z(H\to b\bar{b})$
searches~\cite{Barate:2003sz}
\GW{at a mass of}
about $98 \gev$; due to the $b \bar b$ final state the
mass resolution was rather coarse. The excess observed at LEP can be
expressed in terms of a signal strength as
\begin{equation}
\mu_{bb}^{\rm exp} =\frac{\sigma^{\rm exp}·
\left( e^+e^- \to Z \phi \to Zb\bar{b} \right)}
{\sigma^{\rm SM}\left( e^+e^- \to Z H
    \to Z b \bar b \right)}
= 0.117 \pm 0.057 \; ,
\label{muLEP}
\end{equation}
where in this case the observed cross section times branching ratio is
normalized to the SM expectation for a
SM Higgs boson with a mass of $98\gev$.
The value for $\mu^{\rm exp}_{b b}$ was extracted in~\citere{Cao:2016uwt}
using methods described in~\citere{Azatov:2012bz}.
It should be \GW{noted}
that $\mu_{bb}^{\rm exp}$ was extracted at a slightly
larger mass of $98\gev$ compared to
$\mu_\gamgam^{\rm exp}$ which was extracted assuming a mass of
$95.3\gev$. However, 
\GW{because of the limited mass resolution in the
$\bbbar$ final state at LEP  
the signal strength
of the LEP excess extracted
at $95\gev$ is expected to be} very close
to the value $\mu_{bb}^{\rm exp}$
as stated above, and we therefore use
$\mu_{bb}^{\rm exp}$ obtained at $98\gev$
without modification.

\GW{Since the reported excesses in the $\gamma\gamma$ channel at the LHC 
and the $b\bar{b}$ channel at LEP}
were found at approximately the same mass, 
\GW{the question of a possible common origin received some attention in 
the literature. Specifically it was explored whether certain model 
realizations 
could simultaneously} explain the two excesses, 
while being in agreement with all other Higgs-boson
related limits and measurements.
These possibilities were reviewed
in~\citeres{Heinemeyer:2018jcd,Heinemeyer:2018wzl,Richard:2020jfd}.
Models in which the two excesses can be described
simultaneously
comprise
the extension by a
Higgs singlet with additional vector-like
matter~\cite{Fox:2017uwr},
a radion model~\cite{Richard:2017kot},
type~I 2HDMs  with a moderately-to-strongly fermiophobic CP-even
Higgs boson produced via decays of a charged
Higgs boson lighter than the top quark~\cite{Haisch:2017gql},
a minimal dilaton model~\cite{LiuLiJia:2019kye},
the $\mu\nu$SSM with one~\cite{Biekotter:2017xmf} and three
generations~\cite{Biekotter:2019gtq} of right-handed neutrinos,
a Higgs boson associated with the breakdown of an $U(1)_{L_\mu L_\tau}$
symmetry~\cite{Liu:2018xsw},
a minimum stealth boson model~\cite{AguilarSaavedra:2020wmg},
various realizations of the NMSSM~\cite{Domingo:2018uim,Choi:2019yrv},
including the inflation-inspired $\mu$NMSSM~\cite{Hollik:2018yek},
and the NMSSM with a seesaw extension~\cite{Cao:2019ofo}.
Furthermore, \GW{in \citere{Biekotter:2019kde}
the two excesses were studied} in the 2HDM with an additional real singlet, the
N2HDM~\cite{Chen:2013jvg,Muhlleitner:2016mzt}, with several follow-up
analyses~\cite{Biekotter:2019mib,Biekotter:2019drs,
Biekotter:2020ahz,Biekotter:2020cjs,
Heinemeyer:2021msz}
, where in \citere{Heinemeyer:2021msz} also the
2HDMS (the 2HDM plus a complex singlet and an
additional $Z_3$ symmetry) was analyzed.
In \citere{Biekotter:2021qbc} the possibility of a 
simultaneous description of the two excesses
at 95~GeV and excesses 
reported by ATLAS and CMS near 400~GeV was
investigated in the N2HDM and the NMSSM.
In \citere{Biekotter:2021ovi} a complex singlet
field instead of the real singlet field of the N2HDM
was considered, and as a result also a
valid dark-matter candidate can be present.

\GW{The analysis of \citere{Biekotter:2019kde}}
in the context of
the N2HDM revealed that
only the so-called type~II and type~IV Yukawa structures
can provide \GW{a description} for the diphoton excess observed
at CMS. Here a dominantly singlet-like Higgs boson with
a mass of about $95\gev$ acquires an enhancement of its
branching ratio for the decay into $\gamma\gamma$ by
means of a suppression of the partial decay width for
the $\bbbar$ decay mode.
While comparable values of $\mu_\gamgam$ were shown
to be realized in both the type~II and type~IV
N2HDM, \GW{it was} pointed out in \citere{Biekotter:2019kde}
that the type~IV \GW{scenario} also predicts sizable
branching ratios of the state at $95\gev$ decaying
into pairs of $\tau$-leptons. In contrast, the $\tautau$ decay mode
is suppressed in type~II in the parameter regions
in which the $\gamgam$-excess can be accommodated.
As a consequence, 
\GW{the results for the}
low-mass
Higgs-boson searches in the $\tautau$ final state
\GW{are a crucial test for the N2HDM interpretation of the observed diphoton 
excess which can potentially discriminate between the type~II and type~IV scenarios.}

\GW{In this context the recent results obtained by the CMS collaboration
in the search for additional Higgs bosons in the $\tautau$ channel~\cite{CMS-PAS-HIG-21-001} are obviously of particular interest. 
Remarkably, utilizing the full Run~2 data set, 
in \citere{CMS-PAS-HIG-21-001} the CMS collaboration
reported an excess in the low-mass region for the gluon-fusion production
mode and subsequent decay into $\tautau$ pairs that is compatible with the 
excess that has been observed by CMS in the diphoton search (the latter search 
has not yet been updated to include
the full Run~2 data).}
The excess in the $\tautau$ final state is
most pronounced for a mass hypothesis of
$100\gev$, with a local significance of
$3.1\,\sigma$,
\GW{while for a mass value of $95\gev$, i.e.\ close to the most significant 
excess in the $\gamgam$ search~\cite{Sirunyan:2018aui}, 
CMS reports a local significance of $2.6\,\sigma$.
It should be noted in this context that up to now 
there exists no corresponding 
result for the low-mass search in the $\tautau$ final state from the ATLAS 
collaboration.
For the CMS result,}
the best-fit cross section \GW{for a mass value 
of $95\gev$} has
been determined to be
$\sigma_{gg\phi} \times \mathrm{BR}(\phi \to \tautau) =
(7.7 \pm^{3.9}_{3.1})\,\,$pb~\cite{CMS-PAS-HIG-21-001}.
This corresponds
to a signal strength of
\begin{equation}
\mu^{\rm exp}_{\tau\tau} =
\frac{\sigma\GW{^{\rm exp}}(g g \to \phi \to \tautau)}
{\sigma\GW{^{\rm SM}}(g g \to H \to \tautau)} =
1.2 \pm 0.5 \ ,
\label{muCMSnew}
\end{equation}
where we 
\GW{use a symmetric uncertainty interval for the signal strength that is 
obtained from the lower uncertainty interval of the quoted result for
$\sigma_{gg\phi} \times \mathrm{BR}(\phi \to \tautau)$.}\footnote{The justification of this choice will be given in \refse{sec:numana}.}

The fact that \GW{the LHC searches for a low-mass Higgs boson have 
led to mutually compatible 
excesses in both investigated channels, $\gamgam$ and $\tautau$, 
is a strong motivation for exploring 
a possible BSM nature of the observed patterns. In the present
paper we will focus specifically on the interpretation in the 
context of the N2HDM, as motivated by the earlier analyses in 
\citeres{Biekotter:2019kde,Biekotter:2021qbc}.
\tho{Initially,} we will investigate whether the N2HDM, depending on its Yukawa 
type, can simultaneously describe both the excesses in the 
$\gamgam$ and $\tautau$ channels at the LHC.
\tho{Subsequently,}
we then incorporate also the (statistically slightly less significant)
$\bbbar$ excess observed at LEP into our analysis and investigate 
to what extent the observed patterns in all three search channels
can be successfully described. We will moreover discuss the  
experimental prospects for further probing the possible presence 
of a new state at about $95\gev$ in the near future.}

\GW{The paper is organized as follows.}
After introducing the model \GW{in \refse{sec:n2hdm}}
and the relevant
theoretical and experimental constraints on
the N2HDM parameter space in \GW{\refse{sec:constraints},
the numerical results of our parameter scan are presented in
\refse{sec:numana}, where also the future prospects are discussed.}
We summarize our results in \refse{sec:conclu}.

\section{The N2HDM}
\label{sec:n2hdm}
The N2HDM is the simplest
extension of a CP-conserving Two-Higgs doublet model (2HDM)
in which the latter is augmented with a real
scalar singlet Higgs 
field~\cite{Chen:2013jvg,Muhlleitner:2016mzt}.
\tho{After electroweak symmetry breaking, the fields}
can be parameterized as
\begin{eqnarray}
\Phi_1 = \left( \begin{array}{c} \phi_1^+ \\ \frac{1}{\sqrt{2}} (v_1 +
    \rho_1 + i \eta_1) \end{array} \right) \;, \quad
\Phi_2 = \left( \begin{array}{c} \phi_2^+ \\ \frac{1}{\sqrt{2}} (v_2 +
    \rho_2 + i \eta_2) \end{array} \right) \;, \quad
\Phi_S = v_S + \rho_S \;, \label{eq:n2hdmvevs}
\end{eqnarray}
where $\Phi_1$ and $\Phi_2$ are the two $SU(2)_L$
doublets with hypercharge 1, and
$\Phi_S$ is a real scalar singlet.
The parameters
$v_1, v_2, v_S$ are the real
vacuum expectation values
(vevs) acquired by the fields
$\Phi_1, \Phi_2$ and $\Phi_S$, respectively.
As in the 2HDM we define $\tb := v_2/v_1$.
\tho{A $Z_2$ symmetry is imposed on the
scalar potential,
which is only softly
broken by a bilinear term
usually written as $m_{12}^2 (\Phi_1^\dagger
\Phi_2 + \mathrm{h.c.})$
(see, for instance, Eq.~(2.1) in
\citere{Muhlleitner:2016mzt}). 
The $Z_2$ symmetry is extended to the Yukawa
sector in order to eliminate
tree-level flavor-changing neutral currents.}
As in the 2HDM, one can have four variants of
the N2HDM, depending on the $Z_2$ parities of the
fermions. We will focus on type~II and~IV
(flipped), which were shown to 
be capable of accommodating the diphoton 
excess~\cite{Biekotter:2019kde}.\footnote{In
type~II $\Phi_1$ is coupled to leptons and down-type quarks,
while $\Phi_2$ is coupled to up-type quarks.
In type~IV the couplings to quarks are unchanged,
but the leptons are coupled to $\Phi_2$ instead
of $\Phi_1$.}
In addition, the scalar potential is invariant under a second
$Z_2$ symmetry acting only on $\Phi_S$.
This symmetry is spontaneously broken \GW{if} $\Phi_S$
acquires a vev.

In the CP-even scalar sector,
the states $\rho_1$, $\rho_2$ and
$\rho_S$ mix, leading to a total of three
CP-even physical Higgs bosons
$h_{1,2,3}$, where we use the convention
$\mh1 < \mh2 < \mh3$. The
relation between the two sets of states
is given in
terms of the $3 \times 3$ orthogonal
matrix~$R$, which can be parameterized as
\begin{equation}
\label{mixingmatrix}
R=
\begin{pmatrix}
c_{\alpha_1}c_{\alpha_2} &
  s_{\alpha_1}c_{\alpha_2} &
    s_{\alpha_2} \\
-(c_{\alpha_1}s_{\alpha_2}s_{\alpha_3}+s_{\alpha_1}c_{\alpha_3}) &
  c_{\alpha_1}c_{\alpha_3}-s_{\alpha_1}s_{\alpha_2}s_{\alpha_3}  &
    c_{\alpha_2}s_{\alpha_3} \\
-c_{\alpha_1}s_{\alpha_2}c_{\alpha_3}+s_{\alpha_1}s_{\alpha_3} &
-(c_{\alpha_1}s_{\alpha_3}+s_{\alpha_1}s_{\alpha_2}c_{\alpha_3}) &
c_{\alpha_2}c_{\alpha_3}
\end{pmatrix}~,
\end{equation}
where
$\alpha_1, \alpha_2, \alpha_3$ are the three mixing angles, and
we use the short-hand notations ${s_x = \sin x}$,
${c_x = \cos x}$. The
singlet admixture of the physical states are given by
${\Sigma_{h_i}= |R_{i3}|^2}, {i=1,2,3}$.

The couplings of the Higgs bosons to SM particles are modified
w.r.t.\ to the couplings of
a SM Higgs boson.
We express the couplings of the scalar
mass eigenstates $h_i$, normalized to the corresponding SM couplings,
in terms of the coupling coefficients $c_{h_i V V}$ and
$c_{h_i f \bar f}$, such that the couplings
to the massive vector bosons
are given by
\begin{equation}
\left(g_{h_i W W}\right)_{\mu\nu} =
\mathrm{i} g_{\mu\nu} \left(c_{h_i V V}\right) g M_W
\quad \text{and } \quad
\left(g_{h_i Z Z}\right)_{\mu\nu} =
\mathrm{i} g_{\mu\nu} \left(c_{h_i V V}\right) \frac{g M_Z}{\CW} \, ,
\end{equation}
where $g$ is the $SU(2)_L$ gauge coupling,
$\CW$ is the cosine of the weak
mixing angle, $\CW = \MW/\MZ$, $\SW = \sqrt{1 - \CW^2}$,
and $M_W$ and $M_Z$ are the masses of the $W$ boson
and the $Z$ boson, respectively. The couplings of the Higgs bosons
to the fermions are given by
\begin{equation}
g_{h_i f \bar{f}} =
\frac{\tho{m_f}}{v} \left(c_{h_i f \bar{f}}\right) \; ,
\end{equation}
where $m_f$ is the mass of the fermion, and
$v = \sqrt{(v_1^2 +v_2^2)} \approx 246\gev$ is the SM vev.
Analytical expressions for these coupling coefficients in terms of the
mixing angles $\al_{1,2,3}$ and $\be$ can be found
in~\citere{Biekotter:2019kde}.

The scalar potential of the N2HDM comprises 12
parameters.
Since the value of $v$
can be determined from the known
gauge-boson masses, it can be used to
eliminate one degree of freedom, such
that 11 free parameters remain.
We use the public code \texttt{ScannerS}~\cite{Coimbra:2013qq,
Muhlleitner:2016mzt,Muhlleitner:2020wwk},
\GW{with} which the model
can be explored in terms of the parameters
\begin{equation}
c^{2}_{h_{2}t\bar{t}} \, , \;\;
c^{2}_{h_{2}VV} \, , \;\;
\mathrm{sign}(R_{23}) \, , \;\;
R_{13} \, , \;\;
\tb \, , \;\;
v_{S} \, , \;\;
m_{h_{1,2,3}} \, , \;\;
m_{A} \, , \;\;
\MHp \, , \;\;
m_{12}^{2} \, . \label{eq:inputsnew}
\end{equation}
Here, $m_A$, $\MHp$ denote
the masses of the physical CP-odd and charged Higgs bosons, respectively.
We will identify the lightest CP-even Higgs boson,
$h_1$, with the one
that could
potentially be identified with a signal at
$95 \gev$, labelled \he.
The second-lightest CP-even Higgs boson will be
identified with the observed
state at $125 \gev$, labelled \hz.
Besides the 11 free parameters mentioned above,
\refeq{eq:inputsnew} also contains the entry
$\mathrm{sign}(R_{23})$, which is used
to lift a degeneracy
arising from the dependence of
the mixing angles $\alpha_i$
on the squared values of the coupling
coefficients $c^{2}_{h_{2}t\bar{t}}$
and $c^{2}_{h_{2} VV}$ and the
element of the mixing matrix $R_{13}$.

\section{Theoretical and experimental constraints}
\label{sec:constraints}

\GW{In our analysis we apply several theoretical requirements 
to the parameter space} of the N2HDM.
In order to ensure that the electroweak minimum
of a parameter point is physically viable it
is required that it is either stable or meta-stable,
\GW{where in the latter case}
the electroweak minimum is not the global
minimum of the potential but the
electroweak vacuum is
sufficiently long-lived in comparison to the
age of the universe.
\GW{In particular}, we apply conditions on the scalar couplings
that exclude parameter points for which the
scalar potential is not bounded from
below~\cite{Klimenko:1984qx,Muhlleitner:2016mzt}.
Moreover, for parameter points with a
meta-stable electroweak minimum we calculate the
lifetime of the electroweak vacuum and verify
\GW{that} it is large compared to
the lifetime of the universe. For the calculation of the
lifetime, \texttt{ScannerS} provides an
interface to the public code
\texttt{EVADE}~\cite{Hollik:2018wrr,evade-www} 
\GW{(see also the analysis in \citere{Ferreira:2019iqb})}.
Finally, we apply the
tree-level perturbative unitarity conditions that
ensure that in the high-energy limit
the eigenvalues of the
scalar $2\times 2$ scattering matrix are
smaller than $|8\pi|$~\cite{Muhlleitner:2016mzt}.

The \GW{parameter space} of the N2HDM 
\GW{is also subject to}
various experimental constraints.
We verify the agreement of the \GW{selected} points with
the currently available
measurements of the properties of the state
that has been discovered at
about 125~GeV using the public code
\texttt{HiggsSignals
v.2.6.1}~\cite{Bechtle:2013xfa,Stal:2013hwa,
Bechtle:2014ewa,Bechtle:2020uwn}.
\texttt{HiggsSignals} provides a
statistical $\chi^2$-analysis of the
comparison of the predictions of the considered
model with the measurements of the mass, 
signal strengths \GW{and differential information in terms of STXS bins} of the state at
125~GeV.
In the following we denote as $\chi^2_{125}$
the $\chi^2$ \GW{contribution obtained from} \texttt{HiggsSignals}.
In our scans, we combine the result for
$\chi^2_{125}$ with the $\chi^2$-contribution of the fit result
arising from confronting the predictions for $h_{95}$
with the observed excesses at about 95~GeV,
as we will further specify in \refse{sec:numana}.

In order to test the parameter points
against the exclusion limits from the Higgs-boson
searches at LEP, the
Tevatron and in particular from the LHC,
we employ the public code
\texttt{HiggsBounds v.5.9.1}~\cite{Bechtle:2008jh,Bechtle:2011sb,
Bechtle:2013gu,Bechtle:2013wla,Bechtle:2015pma,Bechtle:2020pkv}.
The limits from
searches for charged Higgs bosons yield important constraints
at low $\tb$~\cite{ATLAS:2020jqj}.
For larger $\tb$ the searches for heavy Higgs bosons decaying
into a pair of $\tau$-leptons play an important role in the type~II
N2HDM~\cite{Sirunyan:2018zut,Aad:2020zxo,CMS-PAS-HIG-21-001}.
The recent CMS \GW{search~\cite{CMS-PAS-HIG-21-001} 
(where the
$\tautau$ excess in the low-mass region has been observed)}
is not yet included in \texttt{HiggsBounds}.
We therefore applied the cross-section limits
\GW{that were obtained} from this search
in the high-mass region in a second step
in addition to the \texttt{HiggsBounds} analysis.
The impact on the allowed parameter points
turned out to be very small, which is related
to the fact that the corresponding ATLAS
analysis \GW{in the high-mass region} 
which is included in \texttt{HiggsBounds}
is more sensitive over a large mass range.
For intermediate values of $\tb$ the
\GW{channels with the highest expected sensitivities arise from} Higgs cascade
decays or bosonic final states including the
massive gauge bosons.

Constraints from flavor-physics observables
are taken into account by
he approach as implemented in \texttt{ScannerS},
where the 2HDM flavor constraints
projected to the $\tan\beta$--$m_{H^\pm}$ plane
as given in \citere{Haller:2018nnx}
are applied under the assumption that the
constrains approximately hold in the N2HDM.
The flavor constraints lead to a
lower limit of $\MHp \tho{\gtrsim} 650\gev$ which is identical
in type~II and type~IV. The lower limit on $\tb$
is somewhat higher in type~IV.

Constraints from electroweak precision observables
(EWPO) can in a simple
approximation be expressed in terms of
the oblique parameters $S$, $T$ and
$U$~\cite{Peskin:1990zt,Peskin:1991sw}.
Effects from physics beyond the SM
on these parameters can be significant
if the new physics contributions
enter mainly through gauge boson self-energies, as it is
the case for extended Higgs sectors.
\texttt{ScannerS} has implemented the one-loop
corrections to the oblique
parameters for models with an arbitrary number of Higgs doublets
and singlets from~\citere{Grimus:2008nb}.
In 2HDMs there is a strong correlation between $T$ and
$U$, and $T$ is the most sensitive of the three oblique parameters.
Thus, the contributions to $U$ are much smaller than the ones to $T$
for points that
are not excluded by an extremely
large value of $T$~\cite{Funk:2011ad},
and can safely be neglected.
Therefore, for points to be in agreement
with the experimental observation, we
require that the predictions for the $S$ and the $T$ parameters
are within the $2 \, \sigma$ ellipse \GW{of the experimental results},
corresponding to
$\chi^2=6.18$ for two degrees of freedom,
making use of the fit result of
\citere{Haller:2018nnx}.

\section{Numerical analysis}
\label{sec:numana}
In this section we discuss our
numerical analysis \GW{where we investigate whether 
the different excesses that were observed near $95\gev$
in the searches for additional Higgs bosons
can be described in terms of a single new Higgs particle.}
In the first part
presented in \refse{sec:numana1}, we restrict
the analysis to the excesses in the
$\gamgam$ and the $\tautau$ final states
observed by CMS.
By comparing the signal rates in
the type~II and the type~IV of
the N2HDM, this analysis will
demonstrate that only in the type~IV N2HDM
the state at $95\gev$ can give rise to
signal rates in the $\gamgam$ and
the $\tau^+\tau^-$ final state
that are large enough to explain the
excesses simultaneously.
In the subsequent analysis
discussed in \refse{sec:numana2},
in which we consequently focus only
on the type~IV, we include
in addition the LEP excess in the $\bbbar$
final state in order to answer the question
whether all three excesses can be
accommodated simultaneously.
We complement the numerical discussion
in \refse{sec:prosp} by investigating
the future prospects for experimentally
confirming or excluding the proposed scenario.
Here we will especially focus on the
differences between the parameter \GW{regions
that describe the two excesses
observed by~CMS and the ones where 
all three excesses can simultaneously be described.}

\subsection{CMS-excesses:
\texorpdfstring{$h_{95} \to \gamma \gamma$}{h95toyy}
and
\texorpdfstring{$h_{95} \to \tau^+ \tau^-$}{h95toll}}
\label{sec:numana1}

\GW{As discussed in 
\citere{Biekotter:2019kde},}
both the type~II
and the type~IV N2HDM can accommodate
the $\gamgam$ excess at $95\gev$.
Here we address the question whether
in addition the $\tautau$ excess can
be explained. To this end, we perform
a scan in \GW{type II and type IV of the N2HDM}
over the free parameters
as defined in \refeq{eq:inputsnew}, where the scan
ranges were chosen to be
\begin{align}
&94 \gev \leq m_{h_1} \leq 98 \gev \; ,
\quad m_{h_2} = 125.09 \gev \; ,
\quad 400 \gev \leq m_{h_3} \leq 1000 \gev \; , \notag \\
&\quad 400 \gev \leq m_A \leq 1000 \gev \; ,
\quad 650 \gev \leq \MHp \leq 1000 \gev \; , \notag\\
&0.5 \leq \tb \leq 14.5 \; ,
\quad 0\gev \leq m_{12}^2 \leq 10^6 \gev^2 \; ,
\quad 100 \gev \leq v_S \leq 1500 \gev \; , \notag \\
&0.6 \leq c_{h_2 VV}^2 \leq 1.0 \; , \quad
0.6 \leq c_{h_2 t \bar t}^2 \leq 1.2 \; , \quad
\mathrm{sign}(R_{13}) = \pm 1 \; , \quad
-1 \leq R_{23} \leq 1 \; . \label{eq:ranges}
\end{align}
For the 
\GW{state $h_1 = h_{95}$
that will be confronted with the observed} excesses,
we use a mass range that is compatible
with the $\gamgam$ excess, 
\GW{since the search in the $\gamgam$ final state} has
the best mass resolution of the three observed excesses.
It should also be noted that we set a
lower limit on $m_{H^\pm}$
\GW{in view of} the constraints from flavor-physics
observables (see above). Since mass splittings between
the states $h_3,A$ and $H^\pm$ larger than
about $200\gev$ give rise to large values
of the quartic scalar couplings and thus
to \GW{potential} problems with pertubation theory or
even Landau poles at energy
scales of around
$1\tev$~\cite{Biekotter:2021ovi}, we accordingly
also limit the scan range of $m_A$ and
$m_{h_3}$ via a lower limit of $400\gev$.
We use the public code
\texttt{ScannerS}~\cite{Coimbra:2013qq,
Muhlleitner:2016mzt,Muhlleitner:2020wwk}, which
scans the parameters randomly over the given
range and applies the theoretical and experimental
constraints discussed in
\refse{sec:constraints}.\footnote{We modified the
routines for the check against the EWPO in order
to \GW{apply} a more conservative two-dimensional 
$\chi^2$-fit to S and T instead of a three-dimensional
$\chi^2$-fit to $S$, $T$ and $U$.}
\texttt{ScannerS} is interfaced to \texttt{HiggsBounds}
and \texttt{HiggsSignals} for the check against
the \GW{limits} from searches for additional
Higgs bosons and the 
\GW{constraints from the measured properties}
of $h_{125}$, respectively. The required theoretical
predictions for the cross sections and the branching
ratios of the scalars are obtained from the public
codes \texttt{SusHi}~\cite{Harlander:2012pb,Harlander:2016hcx}
and \texttt{N2HDECAY}~\cite{Djouadi:1997yw,
Butterworth:2010ym,Djouadi:2018xqq,
Muhlleitner:2016mzt,Engeln:2018mbg}.
These codes also provide the required input
for the computation of the signal rates of
the state $h_{95}$, except for the
Higgsstahlung cross section at LEP, which
we calculate by a rescaling of the SM prediction
with the factor $c_{h_{95}VV}^2$.

\GW{In order to analyze whether a simultaneous fit to 
the observed 
$\gamgam$ and $\tautau$ excesses is possible, 
we} perform a $\chi^2$-analysis
where $\chi^2$ takes
into account the two contributions $\chi^2_{\gamgam}$
and $\chi^2_{\tau\tau}$ defined by the
measured central values
$\mu_{\gamgam,\tau\tau}^{\rm exp}$
and the $1\,\sigma$
uncertainties
$\Delta \mu_{\gamgam,\tau\tau}^{\rm exp}$
of the signal rates related to
\GW{the two excesses as specified} in \refeq{muCMS} and
\refeq{muCMSnew},
i.e.
\begin{equation}
\chi^2_{\gamgam,\tau\tau} =
\frac{
(\mu_{\gamgam,\tau\tau} -
\mu_{\gamgam,\tau\tau}^{\rm exp})^2 }{
(\Delta \mu_{\gamgam,\tau\tau}^{\rm exp})^2} \ ,
\end{equation}
where $\mu_{\gamgam,\tau\tau}$ are the model predictions.
\GW{In view of the fact that in the 
considered extended Higgs sector the properties
of $h_{95}$ are closely related}
to the ones of $h_{125}$, we also
add the contribution $\chi^2_{125}$ obtained
with the help of \texttt{HiggsSignals} in order
to ensure that the properties of
$h_{125}$ are in agreement
with the experimental measurements.
We define
the total $\chi^2$ as the sum 
\begin{equation}
\chi^2 = \chi^2_{\gamgam} +
\chi^2_{\tau\tau} + \chi^2_{125} \ .
\label{eq:nolep}
\end{equation}
In the following we consider a parameter
point as acceptable if the condition
$\chi^2 \leq \chi^2_{\rm SM}$ is fulfilled.
\GW{The $\chi^2$ contribution in the SM,}
$\chi^2_{\rm SM}$, is obtained \GW{for}
$\mu_\gamgam^{\rm SM} = \mu_{\tau\tau}^{\rm SM} = 0$,
yielding $\chi^2_{\mathrm{SM},\gamgam} = 9.00$ and
$\chi^2_{\mathrm{SM},\tau\tau} = 6.17$,
while $\chi^2_{\mathrm{SM},125} = 85.77$
results from \GW{confronting the properties of a 
SM Higgs boson at $125\gev$ with the
experimental measurements using}
\texttt{HiggsSignals}.
We will also indicate the best-fit point
with the smallest value of $\chi^2$ with
a magenta star in our plots.
It should be noted in this context that all the remaining
experimental constraints
are applied on the basis of
approximate 95\% confidence-level
limits, either allowing or excluding a
parameter point, instead of including
additional contributions in the
definition of $\chi^2$ as defined above.
This reflects the fact
that here we are primarily
interested in the collider phenomenology
of the two light states Higgs bosons $h_{95}$
and $h_{125}$.

\begin{figure}
\centering
\includegraphics[width=0.48\textwidth]{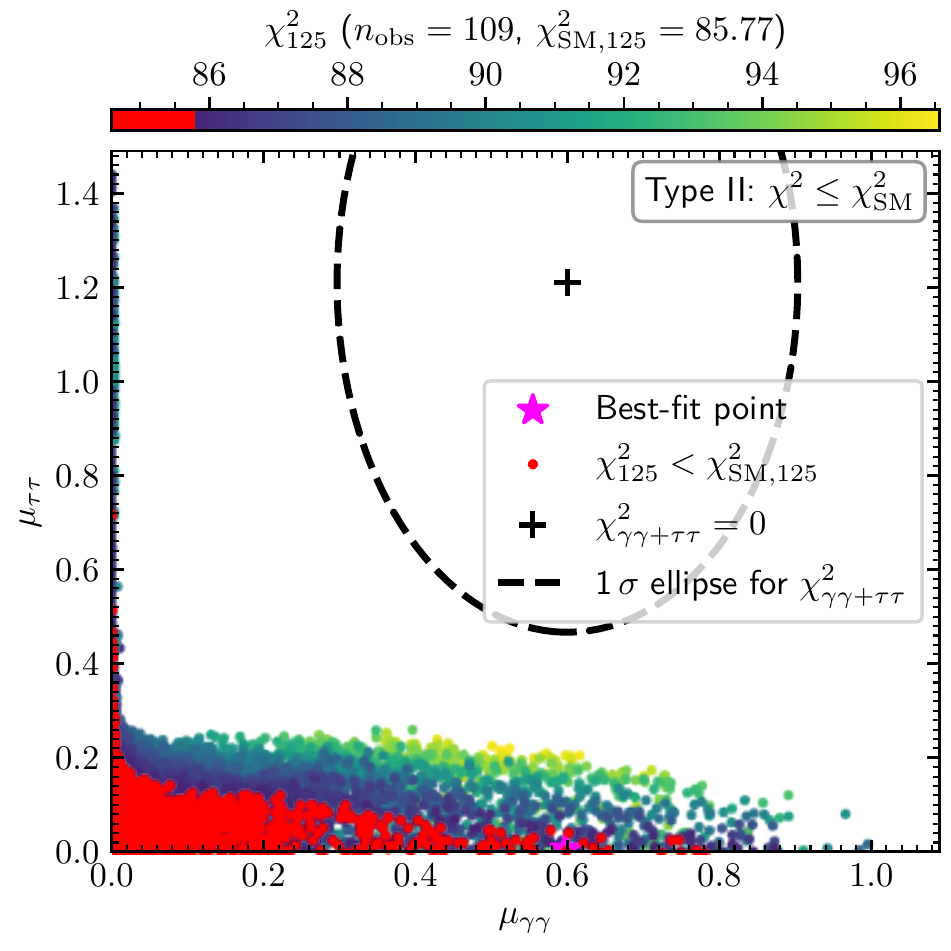}~
\includegraphics[width=0.48\textwidth]{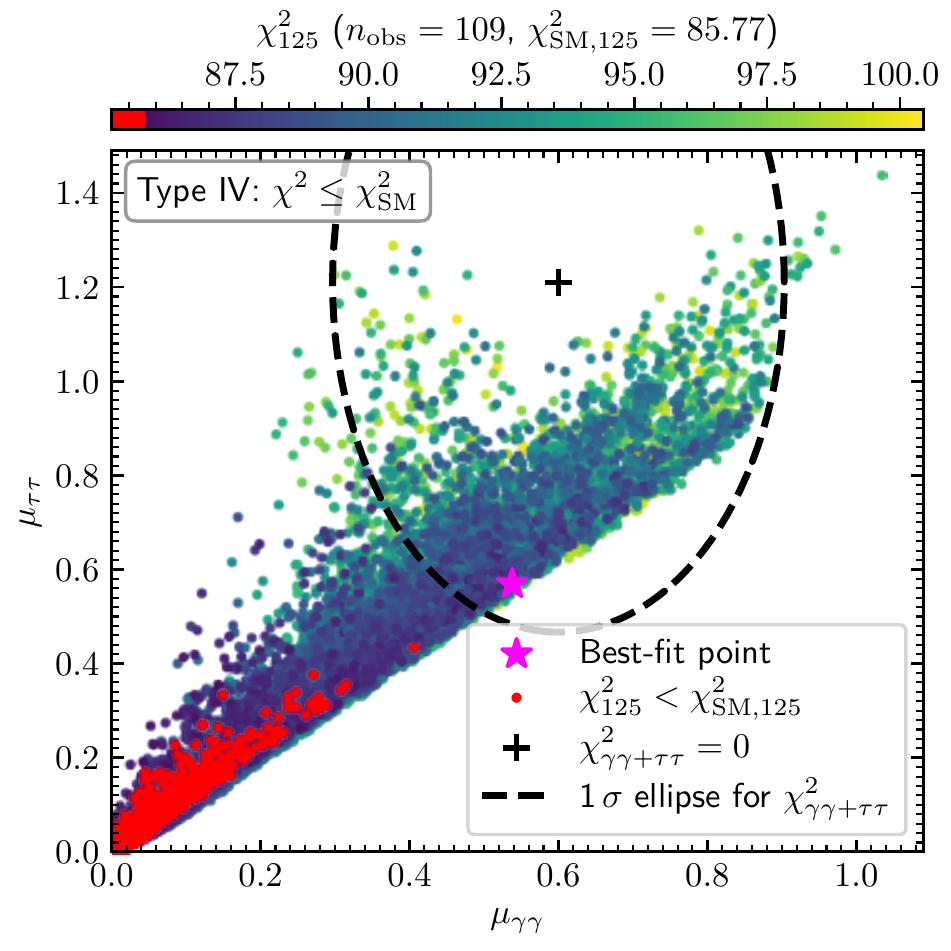}
\caption{\small $\mu_{\tau\tau}$ in dependence
of $\mu_\gamgam$ in the N2HDM type~II (left)
and type~IV (right). The color coding indicates
the value of $\chi^2_{125}$. Red points predict
$\chi^2_{125} < \chi^2_{\mathrm{SM},125}$.
The $1\,\sigma$ confidence-level region with
regards to $\chi^2_{\gamgam+\tau\tau}$
is indicated by the dashed black line.
The best-fit point is indicated with a magenta star.}
\label{fig:muyy_mull_chisqhs_4}
\end{figure}

In \reffi{fig:muyy_mull_chisqhs_4} we show the
parameter points of our scan with the predicted
values for $\mu_\gamgam$ on the horizontal axis
and $\mu_{\tau\tau}$ on the vertical axis.
The colors of the points indicate the value
of $\chi^2_{125}$. In the left plot we show
the parameter points for the type~II N2HDM, and
in the right plot we show the parameter points
for the type~IV.
One can see that only the type~IV N2HDM predicts
parameter points that fall within the $1\,\sigma$
confidence-region
with regards to $\chi^2_{\gamgam+\tau\tau}
= \chi^2_{\gamgam} +
\chi^2_{\tau\tau}$, where the $1\,\sigma$ region
is indicated by the black dashed ellipse 
in \reffi{fig:muyy_mull_chisqhs_4}.\footnote{The
fact that effectively all parameter points
are located below the observed central
value of $\mu_{\tau\tau}^{\rm exp}$ justifies
the choice to use the lower $1\,\sigma$
uncertainty for the cross section in order
to define $\mu^{\rm exp}_{\tau\tau}$
as in \refeq{muCMSnew}.}
On the contrary, parameter points in type~II
that can accommodate the $\tautau$ excess predict
hardly any signal in the $\gamgam$ decay mode,
i.e.~$\mu_{\gamgam} \lesssim 0.02$
if $\mu_{\tau\tau} \gtrsim 0.3$.
As a result, demanding that both excesses are
fitted simultaneously requires a
type~IV interpretation, whereas in type~II
the \GW{two excesses cannot be associated with the same}
particle.\footnote{It remains to be
explored whether supersymmetric models with
additional singlets (e.g.~the NMSSM or the
$\mu\nu$SSM) \GW{could provide a simultaneous} description
of the \GW{two} excesses. In these models potentially
\GW{large} quantum corrections from supersymmetric
partners of the SM particles (e.g.~in terms
of the so-called $\Delta_b$-corrections)
\GW{could} lead to a suppression of the $\bbbar$ decay mode,
while \GW{leaving} the $\tautau$ decay mode
\GW{essentially unsuppressed}.}

The different results in both \GW{Yukawa types 
of the N2HDM} can be
understood by realizing that, as explained
in more detail in \citere{Biekotter:2019kde}, sizable
values of $\mu_\gamgam \approx \mu_\gamgam^{\rm exp}$
require an enhancement of the diphoton
branching ratio via a suppression of the
$h_{95} \to \bbbar$ decay mode.
Accordingly, taking into account that
$c_{h_{95} \bbbar}$ is proportional
to $c_{\alpha_1}$~\cite{Biekotter:2019kde}, fitting
the $\gamgam$ excess is possible in the
\GW{region where
$c_{\alpha_1}$
is small.} In type~II the down-type
quarks \tho{and the leptons}
are coupled to the same Higgs doublet
$\Phi_1$. As a consequence, one finds
$c_{h_{95}\tautau} = c_{h_{95}\bbbar}$, such
that also the $h_{95} \to \tautau$ decay mode
is suppressed in the parameter region suitable
for an explanation of the $\gamgam$ excess.
In type~IV, on the other hand,
the second doublet $\Phi_2$
is coupled to the leptons. Then
$c_{h_{95}\tautau}$ is proportional to
$s_{\alpha_1}$ (instead of
$c_{\alpha_1}$), and the branching ratio
for $h_{95} \to \tautau$ is unsuppressed
in the parameter region that allows
sizable values of $\mu_{\gamgam}$.

In the plots in \reffi{fig:muyy_mull_chisqhs_4}
we have highlighted in red the parameter points that
predict $\Delta \chi^2_{125} =
\chi^2_{125} - \chi^2_{\mathrm{SM},125} < 0$.
Hence, the red points provide \GW{an even} better description
of the measurements of the SM-like Higgs boson
$h_{125}$ than the SM. However, we emphasize that the values of
$\Delta \chi^2_{125}$ for the red points are
so small that they are statistically not
significant. 
\GW{It is interesting to note, however,}
that the red points
all lie outside of the $1\,\sigma$ ellipse with
regards to $\chi^2_{\gamgam} + \chi^2_{\tau\tau}$.
Therefore, small modifications
of the properties of $h_{125}$ compared to the
SM predictions are a feature of the parameter
points that fit both excesses.
This is also the reason why the best-fit
point (magenta star) is located relatively
close to the border of the
$1\,\sigma$ ellipse, because parameter
points located more centrally in the ellipse
are associated with \GW{slightly} larger values of $\chi^2_{125}$.
It should be mentioned, however, that there
are many points in the $1\,\sigma$ ellipse with
$\Delta \chi^2_{125} < 5.99$, which
\GW{would correspond to a 95\% confidence-level
exclusion based just} on the
properties of
$h_{125}$~\cite{Bechtle:2020uwn}.\footnote{Here it
is assumed
that the SM prediction $\chi^2_{\mathrm{SM},125}$
is a good estimate of the best-fit value of the
type~IV N2HDM. This assumption is a good approximation
according to the fact that we found
$\mathrm{min}(\chi^2_{125}) \approx \chi^2_{\mathrm{SM},125}$
in our scans, as is also visible
in the right plot of
\reffi{fig:muyy_mull_chisqhs_4}
and in \reffi{fig:muyy_mull_mubb}.}
Taking into account the current precision of the
signal-rate measurements of $h_{125}$, the modifications
of the theoretical predictions are not large enough
to allow for an exclusion of the points inside
the ellipse.
Nevertheless, future measurements at the HL-LHC
or a possible future $e^+e^-$-collider might be able
to probe the scenario presented here. We will
discuss the future prospects with regards to
the coupling measurements of $h_{125}$ in \GW{more} detail
in \refse{sec:prosp}.

Another interesting observation is that
even in type~IV there
are no points which lie at the center of the
ellipse, which is indicated by the black cross
in \reffi{fig:muyy_mull_chisqhs_4}.
The reason for this is that such points are
excluded by the Higgs-boson searches at
LEP in the di-tau final state~\cite{Schael:2006cr}.
It should be noted in this context that also the points
to the right of the black cross with
$\mu_{\tau\tau} \approx \mu_{\tau\tau}^{\rm exp}$
and $\mu_\gamgam > \mu_{\gamgam}^{\rm exp}$
would be excluded by the LEP search
at the 95\% confidence level.
However, these parameter points still pass
the \texttt{HiggsBounds} analysis
(and are therefore shown in the plots), because
\texttt{HiggsBounds} only compares the predictions
of the cross sections of each Higgs boson
to the most sensitive search based on
the \textit{expected} sensitivity of
the experimental search \GWn{in order to ensure the correct statistical interpretation of the obtained bound as a 95\% C.L. \tho{limit}~\cite{Bechtle:2013wla}.} 
For the parameter points with values
of $\mu_\gamgam > \mu_{\gamgam}^{\rm exp}$
the most sensitive search is the CMS search
in the $\gamgam$ final
state~\cite{Sirunyan:2018aui}, for which
the observed exclusion limit is substantially
weaker than the expected limit due to the
observed excess. \hto{In order to demonstrate the potential impact} \GWn{of the LEP limit in the $\tau^+\tau^-$ final state 
we show in \reffi{fig:referee1} of
App.~\ref{sec:append} the same plot as in \reffi{fig:muyy_mull_chisqhs_4} (right) where we have indicated points that would be excluded 
by the LEP limit in the $\tau^+\tau^-$ final state 
if that channel had been selected in order to 
determine the 95\% C.L.\ limit.
It should be noted, however, that
requiring the limits from several collider searches
to be individually fulfilled at 
the 95\% C.L.\ leads to a
statistical interpretation of the
resulting limit that is stronger than 
an overall 95\% C.L..}
\hto{In order to maintain 
\GWn{the} statistical interpretation of
the applied cross-section limits from
BSM Higgs-boson searches
\GWn{as an overall exclusion bound at the 
95\% C.L., in our analysis} we will
\GWn{stick to} the approach as
implemented in \texttt{HiggsBounds},
\GWn{where} only the} \GWn{observed
limit is applied that has the highest expected
sensitivity.}

\begin{figure}
\centering
\includegraphics[width=0.48\textwidth]{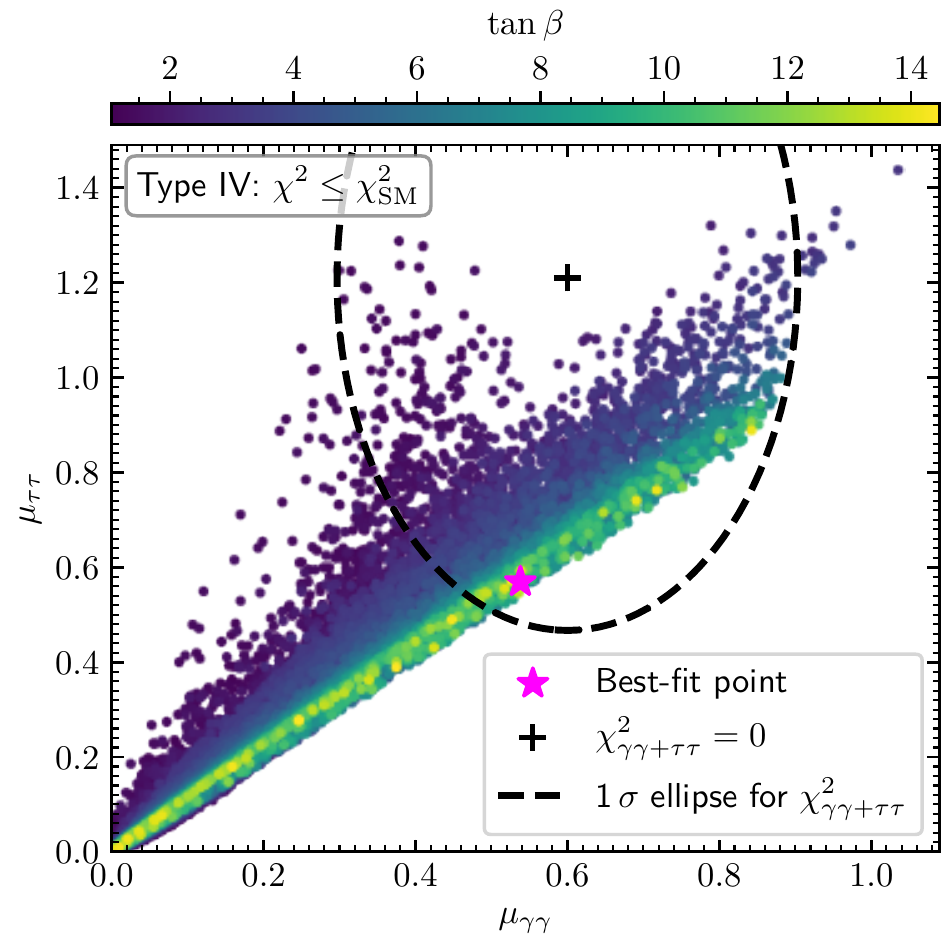}~
\includegraphics[width=0.48\textwidth]{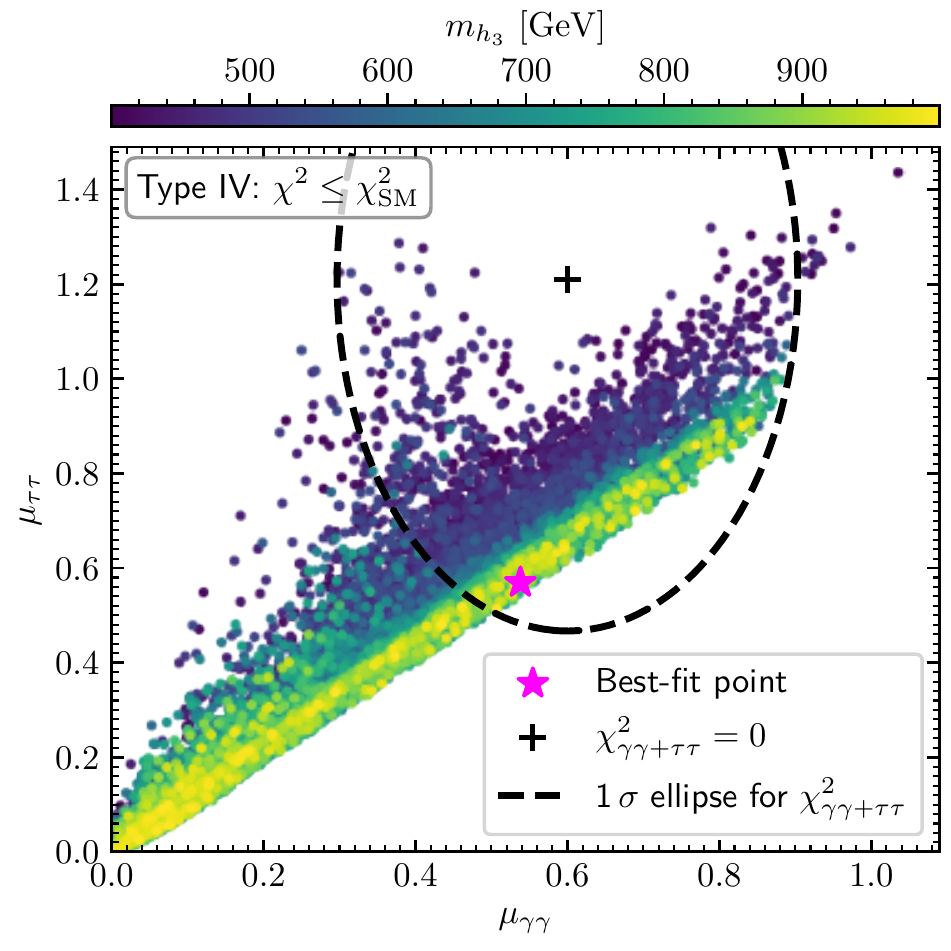}
\caption{\small As in \reffi{fig:muyy_mull_chisqhs_4},
but the color coding indicates
the value of $\tan\beta$ (left) and of $m_{h_3}$ (right).}
\label{fig:muyy_mull_tbetamh3_4}
\end{figure}

After having established that \GW{the type~IV
N2HDM} can account for a simultaneous explanation
of both the $\gamgam$ and the $\tautau$ excesses,
\GW{but not the Yukawa type~II},
we now investigate what parameter configurations
are most suitable. In \reffi{fig:muyy_mull_tbetamh3_4}
we show \GW{for type~IV} 
the parameter points in the same plane
as in \reffi{fig:muyy_mull_chisqhs_4},
but here the color coding indicates
the values of $\tan\beta$ (left plot) and
$m_{h_3}$ (right plot).
Focusing on the points inside the $1\sigma$ ellipse,
one can see that parameter points with
values of $\tan\beta$ at the upper end of the scan range
are located on a diagonal band of points. These points
predict the smallest values of $\mu_{\tau\tau}$
for a given value of $\mu_\gamgam$.
Larger values of $\mu_{\tau\tau}$ for fixed
$\mu_\gamgam$ can be achieved for values
of $\tan\beta \lesssim 5$. In the lower range
of $\tan\beta$ we find parameter points that
reach values of $\mu_{\tau\tau}$ around or
above the observed central value $\mu_{\tau\tau}^{\rm exp}
= 1.2$. No such preference for smaller values
of $\tan\beta$ is observed with regards to
$\mu_\gamgam$.

In the right plot of
\reffi{fig:muyy_mull_tbetamh3_4}
one can observe that the values of $\mu_{\tau\tau}$
are also correlated with the mass of
the heaviest CP-even Higgs boson $m_{h_3}$.
On the diagonal band of points for which
we found the points with large values of
$\tan\beta$, we find values of $m_{h_3}$
over the whole scan range.\footnote{In the right
plot of \reffi{fig:muyy_mull_tbetamh3_4} points
are plotted in ascending order of $m_{h_3}$.}
The parameter points above this diagonal band
that feature the largest values of $\mu_{\tau\tau}$
are only found for values of $m_{h_3}$ at
the lower end of the scan range.
The fact that sizable values of $\mu_{\tau\tau}$
are associated with \GW{relatively} small values of $\tan\beta$
and $m_{h_3}$ is caused by an intricate interplay
of the various theoretical and experimental constrains
which give rise to a correlation between
the allowed values of the mixing angles $\alpha_i$
for given values of $\tan\beta$ and $m_{h_3}$.
The correlations between the parameters
$\tan\beta$ and $m_{h_3}$ and the
signal strength $\mu_{\tau\tau}$
is phenomenologically very
\GW{interesting} in view of the
prospects for collider searches
of the heavy scalar states.
Here it should be noted that in type~IV the prospects
for discovering $h_3$, $A$ and $H^\pm$ in the leptonic
decay modes $h_3,A \to \tautau$ and $H^\pm \to \tau \nu$
are much worse compared to a type~II scenario.
While in type~II for the heavy
states $H$ and $A$ the $\bbbar$-associated
production and the branching ratio
to $\tau\tau$-pairs can be enhanced
with increasing values of $\tan\beta$,
in type~IV the $H,A \to \tautau$ decay
mode is suppressed when the $\bbbar$-associated
production cross section is enhanced.
Consequently, in type~IV
hadronic or bosonic decay modes
play a bigger role.
We will discuss the experimental prospects from
direct searches for the heavy Higgs bosons
in more detail in \refse{sec:prosp}.

\begin{table}
\centering
\small
\begin{tabular}{c c c c c c c c c}
$m_{h_1}$ & $m_{h_2}$ & $m_{h_3}$ & $m_A$ & $\MHp$ & & & & \\
\hline
$
 95.68
$ &
$
125.09
$ &
$
713.24
$ &
$
811.20
$ &
$
677.38
$ & & & & \\
\hline
\hline
$\tb$ & $\alpha_1$ & $\alpha_2$ & $\alpha_3$ & $m_{12}$ & $v_S$ & & & \\
\hline
$
 10.26
$ &
$
  1.57
$ &
$
  1.22
$ &
$
  1.49
$ &
$
221.12
$ &
$
1333.47
$ & & & \\
\hline
\hline
$\br^{bb}_{h_1}$ & $\br^{gg}_{h_1}$ & $\br^{cc}_{h_1}$ &
$\br^{\tau\tau}_{h_1}$ & $\br^{\gamma\gamma}_{h_1}$
& $\br^{WW}_{h_1}$  & $\br^{ZZ}_{h_1}$ & & \\
\hline
$
 0.005
$ &
$
 0.348
$ &
$
 0.198
$ &
$
 0.412
$ &
$
 6.630
\cdot
10^{
-3
}
$ &
$
 0.025
$ &
$
 3.382
\cdot
10^{
-3
}
$ & & \\
\hline
\hline
$\br^{bb}_{h_2}$ & $\br^{gg}_{h_2}$ & $\br^{cc}_{h_2}$ &
$\br^{\tau\tau}_{h_2}$ & $\br^{\ga\ga}_{h_2}$ &
$\br^{WW}_{h_2}$ & $\br^{ZZ}_{h_2}$ & & \\
\hline
$
 0.553
$ &
$
 0.085
$ &
$
 0.032
$ &
$
 0.069
$ &
$
 2.537
\cdot
10^{
-3
}
$ &
$
 0.228
$ &
$
 0.028
$ & & \\
\hline
\hline
$\br^{tt}_{h_3}$ & $\br^{bb}_{h_3}$ & $\br^{\tau\tau}_{h_3}$ & $\br^{h_1 h_1}_{h_3}$ &
$\br^{h_1 h_2}_{h_3}$ & $\br^{h_2 h_2}_{h_3}$ &
$\br^{WW}_{h_3}$ &
 & \\
\hline
$
 0.123
$ &
$
 0.739
$ &
$
 0.000
$ &
$
 0.002
$ &
$
 0.072
$ &
$
 0.030
$ &
$
 0.022
$ &
$

$ &
$

$ \\
\hline
\hline
$\br^{tt}_{A}$ & $\br^{bb}_{A}$ & $\br^{\tau\tau}_{A}$ & $\br^{Z h_1}_{A}$ &
$\br^{Z h_2}_{A}$ &
$\br^{Z h_3}_{A}$ &
$\br^{W \Hpm}_{A}$
& & \\
\hline
$
 0.053
$ &
$
 0.173
$ &
$
 0.000
$ &
$
 0.024
$ &
$
 0.001
$ &
$
 0.015
$ &
$
 0.734
$ & \\
\hline
\hline
$\br^{tb}_{\Hpm}$ & $\br^{\tau\nu}_{\Hpm}$ & $\br^{W h_1}_{\Hpm}$ & $\br^{W h_2}_{\Hpm}$ &
 & & & \\
\hline
$
 0.922
$ &
$
 0.000
$ &
$
 0.073
$ &
$
 0.003
$ &
$

$ &
$

$
& & &
\end{tabular}

\caption{\small Parameters of the best-fit point
for which the minimal value of
$\chi^2$
is found 
($\chi^2 = 88.07$, $\chi^2_{125} = 86.24$)
and branching ratios of the scalar particles in the
type IV scenario.
Dimensionful parameters are given in GeV,
and the angles are given in radian.}
\label{tab:best_4_wobb}
\end{table}

We \GW{complete} the discussion of this section
by a closer examination of the properties
of the best-fit point, which is indicated
by a magenta star in the plots.
The best-fit point has a total $\chi^2$-value
of $\chi^2 = 88.07$, which is composed of
the contributions related to the excesses,
$\chi^2_{\gamgam+\tau\tau} = 1.83$,
and the contribution related to the
SM-like Higgs boson,
$\chi^2_{125} = 86.24$.
Thus, the excesses are described
at the level of less than $1\,\sigma$,
and the properties of $h_{125}$ are practically
indistinguishable from the ones of
a SM Higgs boson given the current
experimental uncertainties.
In \refta{tab:best_4_wobb} we show the
scalar masses and the values of the remaining
free parameters.
One can see that a large branching ratio
for the $\gamgam$ decay mode of $h_{95}$ 
\GW{arises because} $\alpha_1 \approx
\pi / 2$, such that for the coupling
to $b$-quarks one finds $c_{h_{95}\bbbar} \approx 0$.
As a result, also the branching ratio for
$h_{95} \to \bbbar$ is found to be smaller than
1\%. This makes apparent that the best-fit
point of the $\chi^2$-analysis of this section
would not be suitable \GW{for} additionally
accommodating the LEP excess in the $\bbbar$ final state.
Regarding the heavy states, we find that the most
striking collider signature would be associated
to the decay mode $A \to H^\pm W^\mp$. However, given
the \GW{relatively} 
large values of the masses a discovery at the
LHC \GW{seems to be} not very promising.

\subsection{CMS- and LEP-excesses:
\texorpdfstring{$h_{95} \to \gamma \gamma$}{h95toyy},
\texorpdfstring{$h_{95} \to \tau^+ \tau^-$}{h95toll} and
\texorpdfstring{$h_{95} \to b \bar b$}{h95tobb}}
\label{sec:numana2}
As a next step of our analysis, we take into account
also the LEP excess observed
at a comparable mass
in the $\bbbar$ decay mode.
Accordingly, we \GW{investigate whether} 
the singlet-like
scalar $h_{95}$ \GW{in the N2HDM of type~IV can 
have signal-rates that are} in
agreement with the experimentally observed
values in three different decay channels
and two different production modes,
i.e.~gluon-fusion production and
\GW{$e^+e^-$} Higgsstrahlung production.
We restrict the analysis to the type~IV N2HDM
since we demonstrated in the previous section
that the type~II is not capable of accommodating
the $\tautau$ excess in combination
with the $\gamgam$ excess.
We make use of the same set of parameter points
that were generated according to the discussion
in \refse{sec:numana1}. However, \GW{for the present analysis} we
define the total $\chi^2$ that \GW{is investigated} via 
\begin{equation}
\chi^2 = \chi^2_{\gamgam} + \chi^2_{\tau\tau}
+ \chi^2_{bb} + \chi^2_{125} \ ,
\end{equation}
where the additional contribution
\begin{equation}
\chi^2_{bb} =
\frac{
(\mu_{bb} -
\mu_{bb}^{\rm exp})^2 }{
(\Delta \mu_{bb}^{\rm exp})^2} \ ,
\label{eq:yeslep}
\end{equation}
quantifies the description of the $\bbbar$ excess,
constructed
by means of the theory prediction $\mu_{bb}$
and the experimentally
measured central value and $1\,\sigma$ uncertainty
as shown in \refeq{muLEP}.
As before, we consider as valid parameter points
the ones that fulfill the condition
$\chi^2 < \chi^2_{\rm SM}$, where again
$\chi^2_{\rm SM}$ is evaluated assuming no
signal contribution to the excesses at $95\gev$,
such that $\chi^2_{\mathrm{SM},bb} = 4.21$.
The condition $\chi^2 < \chi^2_{\rm SM}$
allows for rather larger values of
$\Delta \chi^2_{125} \gtrsim 15$ for parameter points that
fit all three excesses at
$1\sigma$ or below, i.e.~$\chi^2_{\gamgam+
\tau\tau + bb} = \chi^2_{\gamgam} + \chi^2_{\tau\tau}
+ \chi^2_{bb} < 3.53$.
In order to ensure that there are parameter
points that fit the excesses and which are not
significantly disfavoured by the experimental
data related to $h_{125}$,
we will therefore also show results under the additional
constraint $\Delta \chi^2_{125} < 5.99$.

\begin{figure}
\centering
\includegraphics[width=0.48\textwidth]{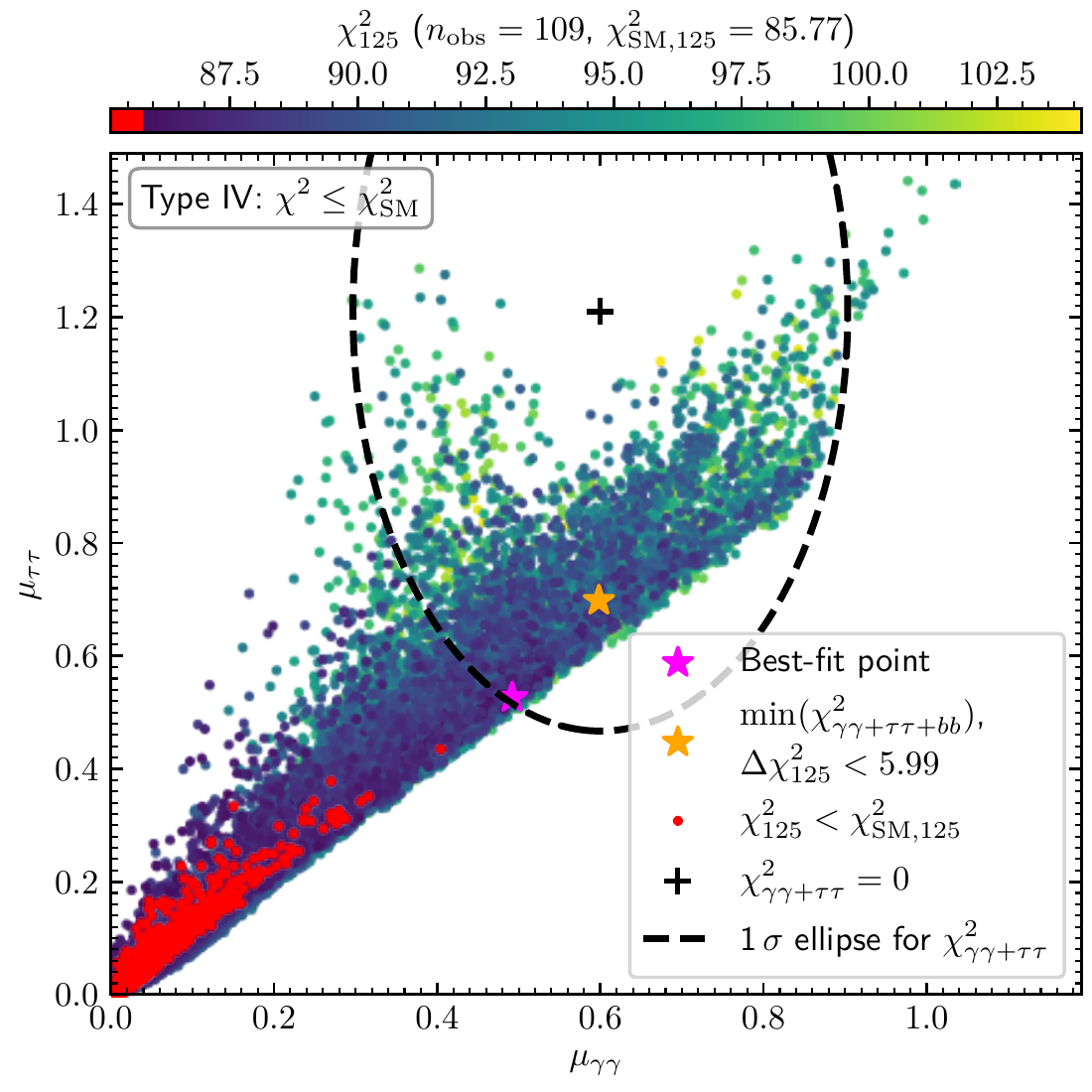}~
\includegraphics[width=0.48\textwidth]{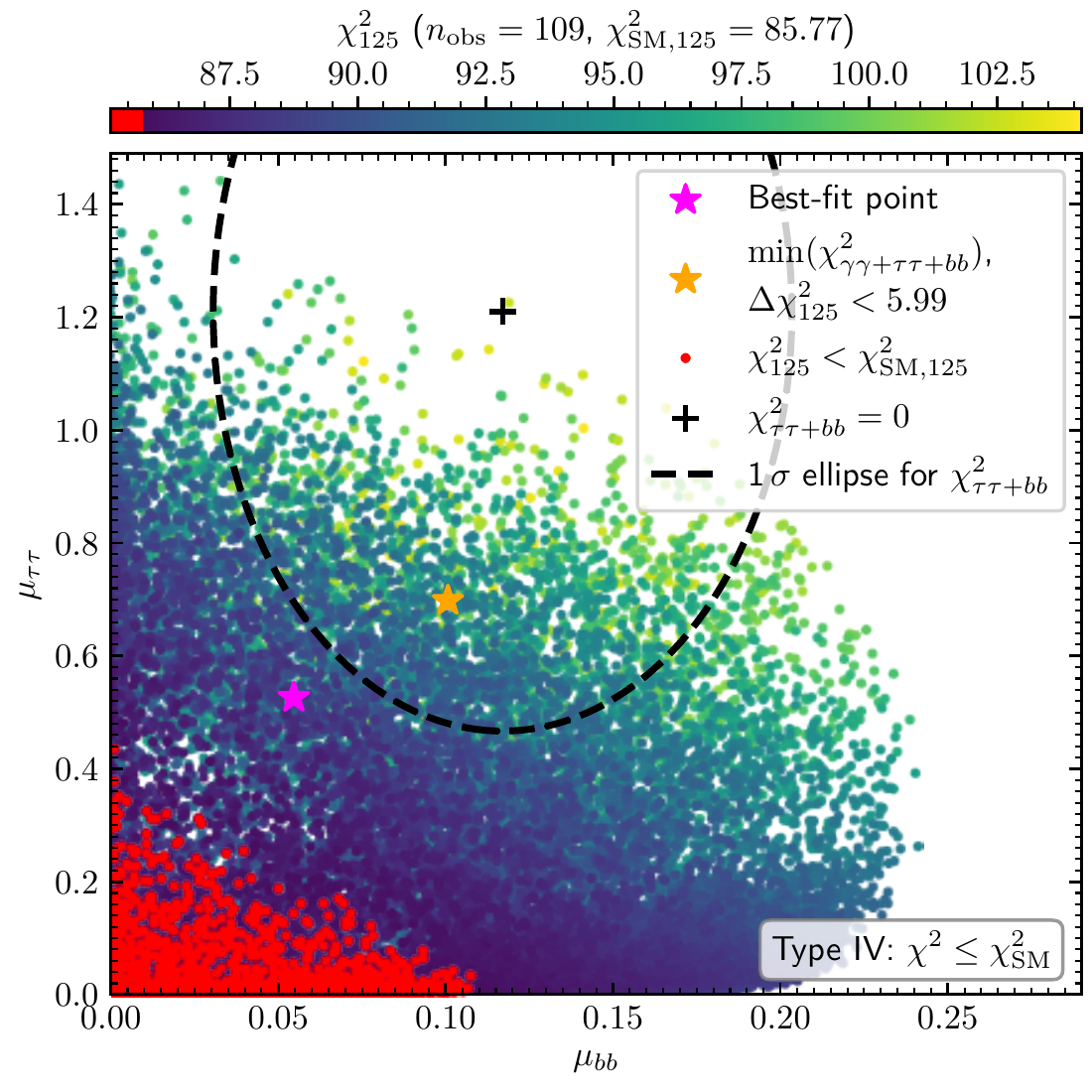}
\caption{\small $\mu_{\tau\tau}$ in dependence of
$\mu_\gamgam$ (left) and $\mu_{bb}$ (right).
The $1\,\sigma$ confidence-level regions with
regard to $\chi^2_\gamgam + \chi^2_{\tau\tau}$ (left)
and $\chi^2_{bb} + \chi^2_{\tau\tau}$ (right)
are indicated by the dashed black line.
The color coding is as in \reffi{fig:muyy_mull_chisqhs_4}.}
\label{fig:muyy_mull_mubb}
\end{figure}

In the left plot of \reffi{fig:muyy_mull_mubb}
we show the parameter points in the
$\mu_\gamgam$--$\mu_{\tau\tau}$ plane, with the
color coding indicating the value of $\chi^2_{125}$.
This plot can be compared to the right plot
of \reffi{fig:muyy_mull_chisqhs_4} from the
analysis discussed in \refse{sec:numana1}.
One can see that the distribution of the points
is very similar in both plots. This indicates
that the inclusion of $\chi^2_{bb}$ has no
significant impact on the values of
$\mu_\gamgam$ and $\mu_{\tau\tau}$ that
can be achieved while respecting the
condition $\chi^2 < \chi^2_{\rm SM}$.
However, the best-fit point, indicated
with the magenta star, has changed, and it
is located further away from the center
of the $1\,\sigma$ ellipse.
This indicates that the parameter points
that have a better agreement with the
observed signal rates of the excesses
are associated with a $\chi^2$-penalty
from the properties of $h_{125}$ contained
in $\chi^2_{125}$. In order to shed light on
whether it is possible to describe the
excesses sufficiently well without being
in significant tension with the signal-rate
measurements of $h_{125}$, we indicate
with the orange star the parameter point with
the minimum value of $\chi^2_{\gamgam+\tau\tau+bb} =
\chi^2_\gamgam + \chi^2_{\tau\tau} + \chi^2_{bb}$
while additionally fulfilling the
condition $\Delta \chi^2_{125} < 5.99$.
One can see that the orange star is located
within the $1\,\sigma$ ellipse regarding
the two-dimensional $\chi^2$-distribution
$\chi^2_{\gamgam+\tau\tau} =
\chi^2_\gamgam + \chi^2_{\tau\tau}$, indicating
a good description of both excesses. 
\hto{As \GWn{for the case of} the right plot of \reffi{fig:muyy_mull_chisqhs_4}, 
we show in the left plot of \reffi{fig:referee2} in 
App.~\ref{sec:append}} \GWn{the same plot as in
\reffi{fig:muyy_mull_mubb} (left) where we have indicated in grey points that would be excluded 
by the LEP limit in the $\tau^+\tau^-$ final state 
if that channel had been selected in order to 
determine the 95\% C.L.\ limit.}

In the right plot of
\reffi{fig:muyy_mull_mubb} we show the
parameter points in the $\mu_{bb}$--$\mu_{\tau\tau}$
plane, with the same color coding as
in the left plot.
The black ellipse in the right plot
shows the $1\,\sigma$ region with regards to
$\chi^2_{bb+\tau\tau} =
\chi^2_{bb} + \chi^2_{\tau\tau}$.
\GW{One can} see that many points lie within the
$1\sigma$ ellipse. Thus, both the $\bbbar$
excess and the $\tautau$ excess can be
accommodated simultaneously. Moreover,
the orange star, defined as described above,
lies within the ellipse. Since the orange
star also lies within the ellipse in the
left plot of \reffi{fig:muyy_mull_mubb}
one can conclude that there are points
which give rise to a good description
of all three excesses, without being
in tension with the experimental data
related to the SM-like Higgs boson $h_{125}$.
However, the best-fit point lies outside
of the $1\sigma$ ellipse in the
right plot of \reffi{fig:muyy_mull_mubb}. Accordingly,
it can also be observed that the
values of $\chi^2_{125}$ grow with increasing
values of the signal-strength
$\mu_{bb}$ and $\mu_{\tautau}$.
This tendency is mainly related to
the couplings of $h_{95}$ and $h_{125}$
to vector bosons, which fulfill the
sum rule $c_{h_{95}VV}^2 + c_{h_{125}VV}^2 \approx 1$
taking into account that the vector-boson
coupling of the heavy state $h_3$ is
negligible. In order to account
for the LEP excess, the Higgsstrahlung
production is sufficiently large if
$c_{h_{95}VV}^2 \approx 0.117 / \mathrm{BR}_{h_{95}}^{bb}$
(see \refeq{muLEP}).
This means that depending on the branching
ratio for $h_{95} \to \bbbar$ the value
of $c_{h_{125}VV}^2$ is suppressed by at least 10\%.
Since the $h_{95} \to \tautau$ decay mode
competes with the decay into $\bbbar$,
a sizable value of $\mu_{\tau\tau}$ requires
an even larger value of $c_{h_{95}VV}^2$
in order to fit the $\bbbar$ excess, which
further suppresses the value of $c_{h_{125}VV}^2$
and strengthens the tension with the measured
signal rates of $h_{125}$.
Hence, we expect to observe larger modifications
of the properties of $h_{125}$ compared to the
SM predictions in this scenario compared to
the previous scenario discussed in \refse{sec:numana1}
in which the $\bbbar$ excess was not considered.
\hto{As for the \GWn{case of the} 
left plot of this figure,  
we show in the right plot of \reffi{fig:referee2} in 
App.~\ref{sec:append}}
\GWn{the same plot as in
\reffi{fig:muyy_mull_mubb} (right) where we have indicated in grey points that would be excluded 
by the LEP limit in the $\tau^+\tau^-$ final state 
if that channel had been selected in order to 
determine the 95\% C.L.\ limit.}

\begin{figure}[t]
\centering
\includegraphics[width=0.84\textwidth]{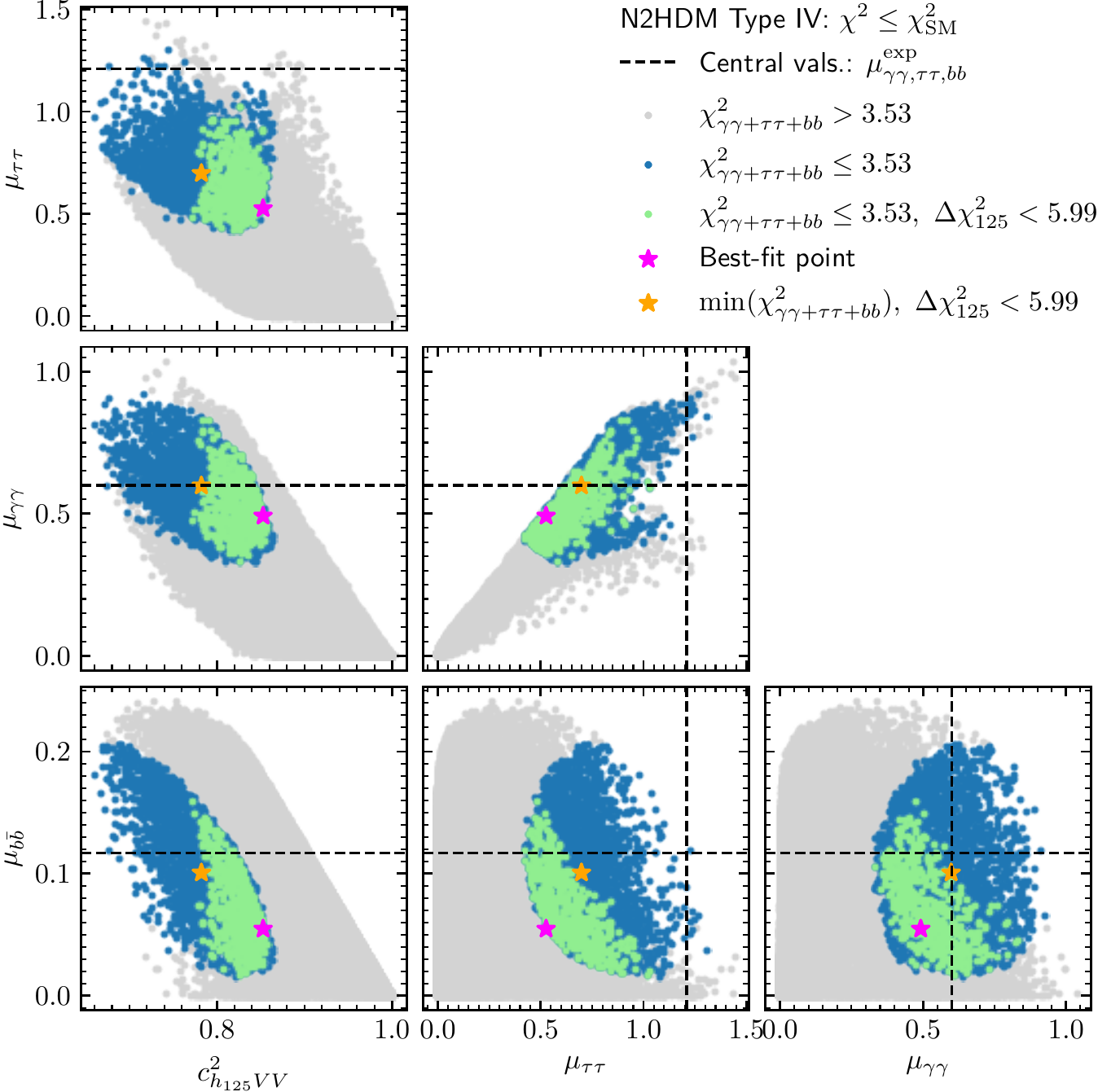}
\caption{\small Correlations between the signal rates
$\mu_{\tau\tau}$, $\mu_{\gamgam}$ and
$\mu_{\bbbar}$, and the coupling coefficient
$c_{h_{125}VV}^2$. Parameter points that fit
the excesses within a \GW{three-dimensional}
confidence level of
$1\,\sigma$ are shown in blue, whereas the
remaining points are shown in 
\GWn{grey.}
Green points fit the excesses within a
confidence level of $1\,\sigma$ and additionally
fulfill the condition $\Delta \chi^2_{125} < 5.99$.
The best-fit point ($\chi^2 = 90.86$,
$\chi^2_{125} = 87.40$) is indicated with a
magenta star.
The orange star ($\chi^2 = 92.66$,
$\chi^2_{125} = 91.46$) indicates the point
with the minimal value of
$\chi^2_{\gamgam+\tau\tau+bb}$
under the condition that
$\Delta \chi^2_{125} < 5.99$.}
\label{fig:mus_main}
\end{figure}

To further scrutinize the relations
between the signal rates of the state
$h_{95}$ among themselves
and also between the former
and the properties
of the state $h_{125}$, we show in
\reffi{fig:mus_main} the correlations of
$\mu_\gamgam$, $\mu_{\tau\tau}$, $\mu_{bb}$
and $c_{h_{125}VV}^2$.
In these plots the parameter points are
shown in three different colors. The 
\GWn{grey}
points are points that do not fit the
excesses inside of the three-dimensional
$1\,\sigma$ ellipsoid corresponding
to $\chi^2_{\gamgam+\tau\tau+bb} \leq 3.53$
and are therefore not
further discussed in the following.
The blue points are points inside the
$1\,\sigma$ preferred regions according to
the observed values of $\mu_\gamgam$,
$\mu_{\tau\tau}$ and $\mu_{bb}$,
i.e.~they feature $\chi^2_{\gamgam+\tau\tau+bb} < 3.53$.
Finally, the green points are the subset
of the blue points that additionally
fulfill the condition $\Delta \chi^2_{125} < 5.99$,
and as such they do not feature
\GW{large} modifications
of the signal rates of $h_{125}$ 
\GW{in view of}
the current experimental uncertainties.
One can see that for all three excesses
there are blue points that reach the experimentally
observed central values of the signal strengths.
However, the experimental central value of
$\mu_{\tau\tau}$
cannot be reached when the additional constraint on
$\Delta \chi^2_{125}$ is considered,
as can bee seen in the plot in the first row.
In the lower right plot one can furthermore
see that the central values of both $\mu_\gamgam$
and $\mu_{bb}$ can be reached simultaneously.
In the lower center plot, on the other hand,
there are blue points that reproduce the
central values of $\mu_{\tau\tau}$ and
$\mu_{bb}$, but the green points
feature smaller values of either $\mu_{\tau\tau}$
or $\mu_{bb}$. Finally, in the right plot in
the second row there are both blue and green
points in the vicinity of the central values
of $\mu_{\tau\tau}$ and $\mu_{\gamgam}$.
However, no points are found exactly at the
central values due to the exclusion
limits of the LEP search $ee \to Z (h \to \tautau)$,
as \GW{already} discussed 
in \refse{sec:numana1}.

Regarding the values of $c_{h_{125}VV}^2$ that
are preferred by the excesses, one can see
that only for
$c_{h_{125}VV}^2 \lesssim 0.86$ there are
points that fit the excesses sufficiently well.
While there are blue points in the range
$0.66 \lesssim c_{h_{125}VV}^2 \lesssim 0.86$,
the green points stretch over a substantially
reduced interval of
$0.76 \lesssim c_{h_{125}VV}^2 \lesssim 0.86$.
It is also interesting to compare the maximum
value of $c_{h_{125}VV}^2$ for which each excess
on its own can be accommodated, and how these numbers
compare to the case where all three excesses are
fitted together. One can see by comparing the
plots in the first column that values of
$c_{h_{125}VV}^2 \approx 0.90$ are sufficiently
small in order to fit the central value
of the signal strength of 
the $\bbbar$ excess individually, and a somewhat
smaller value of $c_{h_{125}VV}^2 \approx 0.88$
is required for the central value of the
$\gamgam$ excess.
These values are slightly above the range of
$c_{h_{125}VV}^2$ for which the excesses can
be fitted simultaneously.
The \GW{relatively} large modifications of the couplings
of $h_{125}$ compared to the SM prediction
are a clear collider target that can be \GW{explored}
at the high-luminosity phase of
the LHC or at possible future lepton colliders,
as will be discussed in more detail
in \refse{sec:prosp}.

\begin{figure}
\centering
\includegraphics[width=0.48\textwidth]{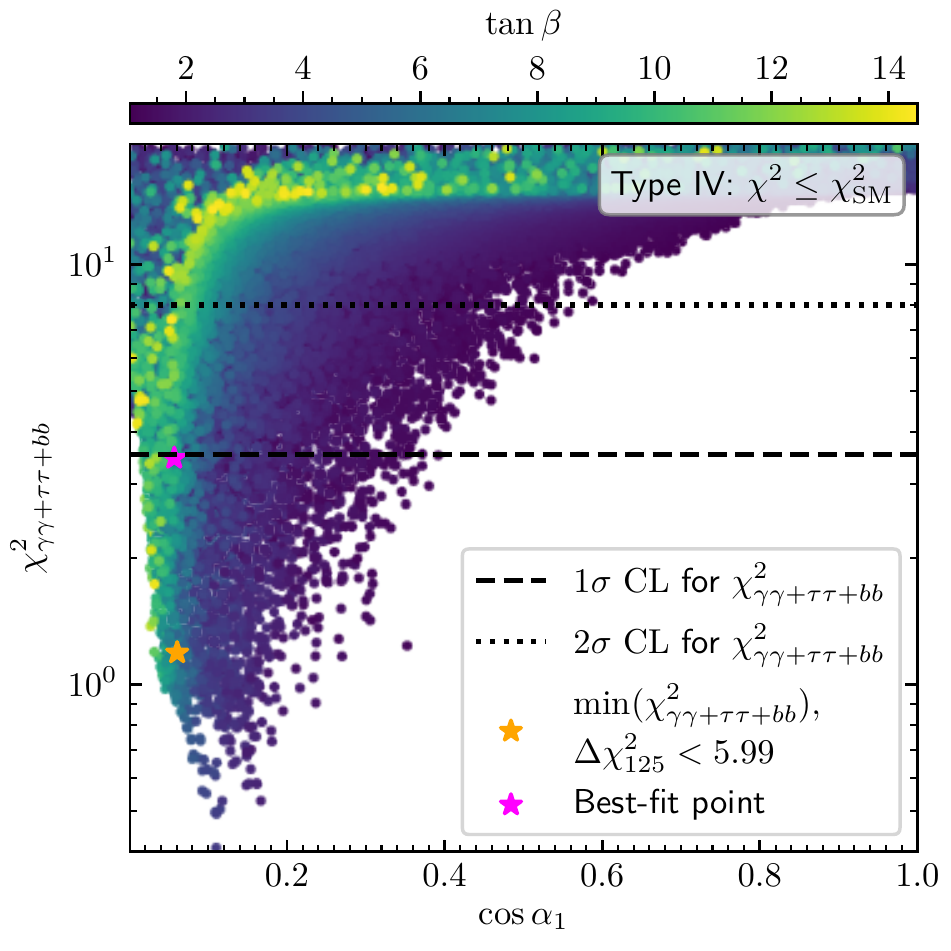}~
\includegraphics[width=0.48\textwidth]{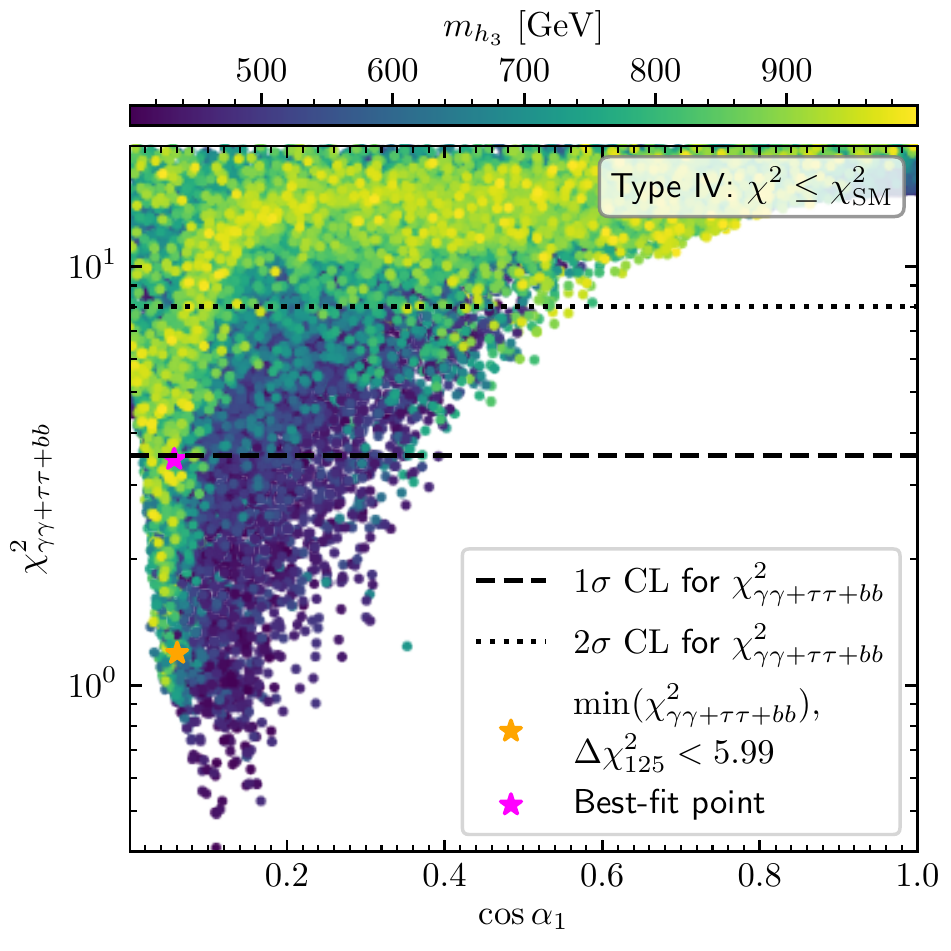}
\caption{$\chi^2_{\gamgam+\tau\tau+bb}$ in dependence
of $\cos\alpha_1$. The color coding indicates the
values of $\tan\beta$ (left) and $m_{h_3}$ (right).
The magenta and orange stars are defined as
in \reffi{fig:muyy_mull_mubb}. The horizontal
dashed and dotted lines indicate the $1\,\sigma$
and the $2\,\sigma$ regions, respectively.}
\label{fig:chisq}
\end{figure}

We \GW{finish} this section
with a discussion of the values of
$\chi^2_{\gamgam+\tau\tau+bb}$ that we
found in our scan and \GW{of the} parameter
region \GW{where} the best description of the excesses
has been achieved.
In \reffi{fig:chisq} we show the
distribution of $\chi^2_{\gamgam+\tau\tau+bb}$
in dependence of $\cos\alpha_1$.
\GW{As discussed in \citere{Biekotter:2019kde}
(and also in \refse{sec:numana1} above)},
both the $\gamgam$ excess and the $\tautau$
excess require small values of
$\cos\alpha_1$ in order to enhance the
branching ratios of the decay modes
$h_{95} \to \gamgam,\tautau$.
For that reason, we find values of
$\chi^2_{\gamgam+\tau\tau+bb} < 3.53$,
corresponding to the $1\,\sigma$ region,
at the lower range of $\cos\alpha_1$.
However, for the smallest values of
$\cos\alpha_1$ the values of $\chi^2_{\gamgam+\tau\tau+bb}$
increase drastically. The reason is that
here the decay mode $h_{95} \to \bbbar$
has a tiny branching ratio, and the $\bbbar$
excess is therefore not accounted for.
In the range $\cos\alpha_1 \gtrsim 0.1$
we only find parameter points with
values of $\tan\beta$ at the lower end of
the scan range.
These are also the points which have the
smallest value of $\chi^2_{\gamgam+\tau\tau+bb}$,
\GW{with a} minimum at $\chi^2_{\gamgam+\tau\tau+bb}
\approx 0.5$.
The orange star,
indicating the parameter point with minimal
value of $\chi^2_{\gamgam+\tau\tau+bb}$
while fulfilling $\Delta\chi^2_{125} < 5.99$,
is located well below the $1\,\sigma$
\GW{level} $\chi^2_{\gamgam+\tau\tau+bb} = 3.53$.
In the right plot of \reffi{fig:chisq}
the color coding indicates the value
of $m_{h_3}$. One can see a similar
correlation between $\cos\alpha_1$
and $m_{h_3}$ as in the left plot
for $\tan\beta$: Only parameter points
for which $m_{h_3}$ has a value at the
lower end of the scan range lie below
the $1\,\sigma$ \GW{level} for $\cos\alpha_1
\gtrsim 0.1$. On the other hand,
for smaller values of
$\cos\alpha_1$, there are points
covering the whole scan range of
$m_{h_3}$ in the $1\,\sigma$ region, where
larger values of $m_{h_3}$ correlate
with larger values of $\tan\beta$.

\begin{table}
\centering
\small
\begin{tabular}{c c c c c c c c c}
$m_{h_1}$ & $m_{h_2}$ & $m_{h_3}$ & $m_A$ & $\MHp$ & & & & \\
\hline
$
 96.47
$ &
$
125.09
$ &
$
733.28
$ &
$
705.87
$ &
$
776.61
$ & & & & \\
\hline
\hline
$\tb$ & $\alpha_1$ & $\alpha_2$ & $\alpha_3$ & $m_{12}$ & $v_S$ & & & \\
\hline
$
  6.54
$ &
$
 -1.51
$ &
$
 -1.08
$ &
$
 -1.41
$ &
$
283.92
$ &
$
1244.58
$ & & & \\
\hline
\hline
$\br^{bb}_{h_1}$ & $\br^{gg}_{h_1}$ & $\br^{cc}_{h_1}$ &
$\br^{\tau\tau}_{h_1}$ & $\br^{\gamma\gamma}_{h_1}$
& $\br^{WW}_{h_1}$  & $\br^{ZZ}_{h_1}$ & & \\
\hline
$
 0.377
$ &
$
 0.230
$ &
$
 0.120
$ &
$
 0.250
$ &
$
 3.714
\cdot
10^{
-3
}
$ &
$
 0.016
$ &
$
 2.116
\cdot
10^{
-3
}
$ & & \\
\hline
\hline
$\br^{bb}_{h_2}$ & $\br^{gg}_{h_2}$ & $\br^{cc}_{h_2}$ &
$\br^{\tau\tau}_{h_2}$ & $\br^{\ga\ga}_{h_2}$ &
$\br^{WW}_{h_2}$ & $\br^{ZZ}_{h_2}$ & & \\
\hline
$
 0.489
$ &
$
 0.099
$ &
$
 0.036
$ &
$
 0.079
$ &
$
 2.808
\cdot
10^{
-3
}
$ &
$
 0.259
$ &
$
 0.032
$ & & \\
\hline
\hline
$\br^{tt}_{h_3}$ & $\br^{bb}_{h_3}$ & $\br^{\tau\tau}_{h_3}$ & $\br^{h_1 h_1}_{h_3}$ &
$\br^{h_1 h_2}_{h_3}$ & $\br^{h_2 h_2}_{h_3}$ &
$\br^{WW}_{h_3}$ &
 & \\
\hline
$
 0.067
$ &
$
 0.150
$ &
$
 0.000
$ &
$
 0.002
$ &
$
 0.318
$ &
$
 0.182
$ &
$
 0.188
$ &
$

$ &
$

$ \\
\hline
\hline
$\br^{tt}_{A}$ & $\br^{bb}_{A}$ & $\br^{\tau\tau}_{A}$ & $\br^{Z h_1}_{A}$ &
$\br^{Z h_2}_{A}$ &
 & & & \\
\hline
$
 0.337
$ &
$
 0.185
$ &
$
 0.000
$ &
$
 0.450
$ &
$
 0.027
$ &
$

$ &
$

$ & \\
\hline
\hline
$\br^{tb}_{\Hpm}$ & $\br^{\tau\nu}_{\Hpm}$ & $\br^{W h_1}_{\Hpm}$ & $\br^{W h_2}_{\Hpm}$ &
 & & & \\
\hline
$
 0.450
$ &
$
 0.000
$ &
$
 0.516
$ &
$
 0.031
$ &
$

$ &
$

$
& & &
\end{tabular}

\caption{\small Parameters
for the point
for which the minimal value of
$\chi^2_{\gamgam+\tau\tau+bb}$
is found under the condition that
$\Delta \chi^2_{125} < 5.99$ (orange star,
$\chi^2 = 92.66$,
$\chi^2_{125} = 91.47$).
Also shown are the
branching ratios of the scalar particles.
Dimensionful parameters are given in GeV,
and the angles are given in radian.}
\label{tab:orange}
\end{table}

\GW{In order to provide a concrete}
example of
a parameter point that fits the
three excesses,
we show in
\refta{tab:orange} the
spectrum and the other free parameters
of the parameter point indicated
with the orange star in the plots.
This parameter point predicts
\begin{equation}
{\color{orange}\bigstar}
\quad
\mu_\gamgam = 0.60
\quad
\mu_{\tau\tau} = 0.70
\quad
\mu_{bb} = 0.10
\ .
\end{equation}
This results in a total $\chi^2$-value
of $\chi^2 = 92.66$, where
$\chi^2_{125} = 91.47$.
It is interesting to compare the
branching ratios of $h_{95}$ to the
ones that we found for the best-fit
point in the analysis discussed in
\refse{sec:numana1} in which the $\bbbar$
excess was not taken into account.
In \GW{that case we found a best-fit point for which the} branching ratio for
the decay mode $h_{95} \to \bbbar$ is
vanishing due to $\alpha_1 \approx \pi / 2$,
whereas in the parameter point shown
in \refta{tab:orange} the branching ratio
for the decay into $\bbbar$ is still sizable
since $|\alpha_1|$ is slightly smaller.
Another important difference concerns the
branching ratios of the heavier states.
While for the best-fit point from \refse{sec:numana1}
the branching ratios for the decays of
$h_3$, $A$ and $H^\pm$ into final states
including $h_{95}$ are small, here we find that
$A$ and $H^\pm$ decay with roughly
50\% probability into $h_{95}$ plus
a gauge boson, and $h_3$ decays with
a probability of about 50\% into
$h_{95}h_{125}$- or $h_{125}h_{125}$-pairs.
These decay signatures are \GW{potentially} 
accessible and
already searched for at the
LHC~\cite{CMS:2016xnc,ATLAS:2017xel,ATLAS:2018oht,
CMS:2019ogx,CMS:2019qcx,
ATLAS-CONF-2020-043,CMS:2021yci}.
\GW{These searches therefore offer good prospects for 
future tests of the considered scenario}
with a particle state at $95\gev$
that gives a good description
of the excesses.

\subsection{Future prospects}
\label{sec:prosp}
In our previous discussion we already
touched upon ways to indirectly or
directly test the existence
of the hypothesized state at $95\gev$.
In this section we give concrete
examples as to where deviations from the
SM predictions might show up in
future collider experiments.
We start by discussing the modifications
of the couplings of the SM-like Higgs boson
at $125\gev$. Afterwards, we discuss how
the scenarios discussed above can be probed
via direct searches for BSM Higgs bosons.

\begin{figure}[t]
\centering
\includegraphics[width=0.48\textwidth]{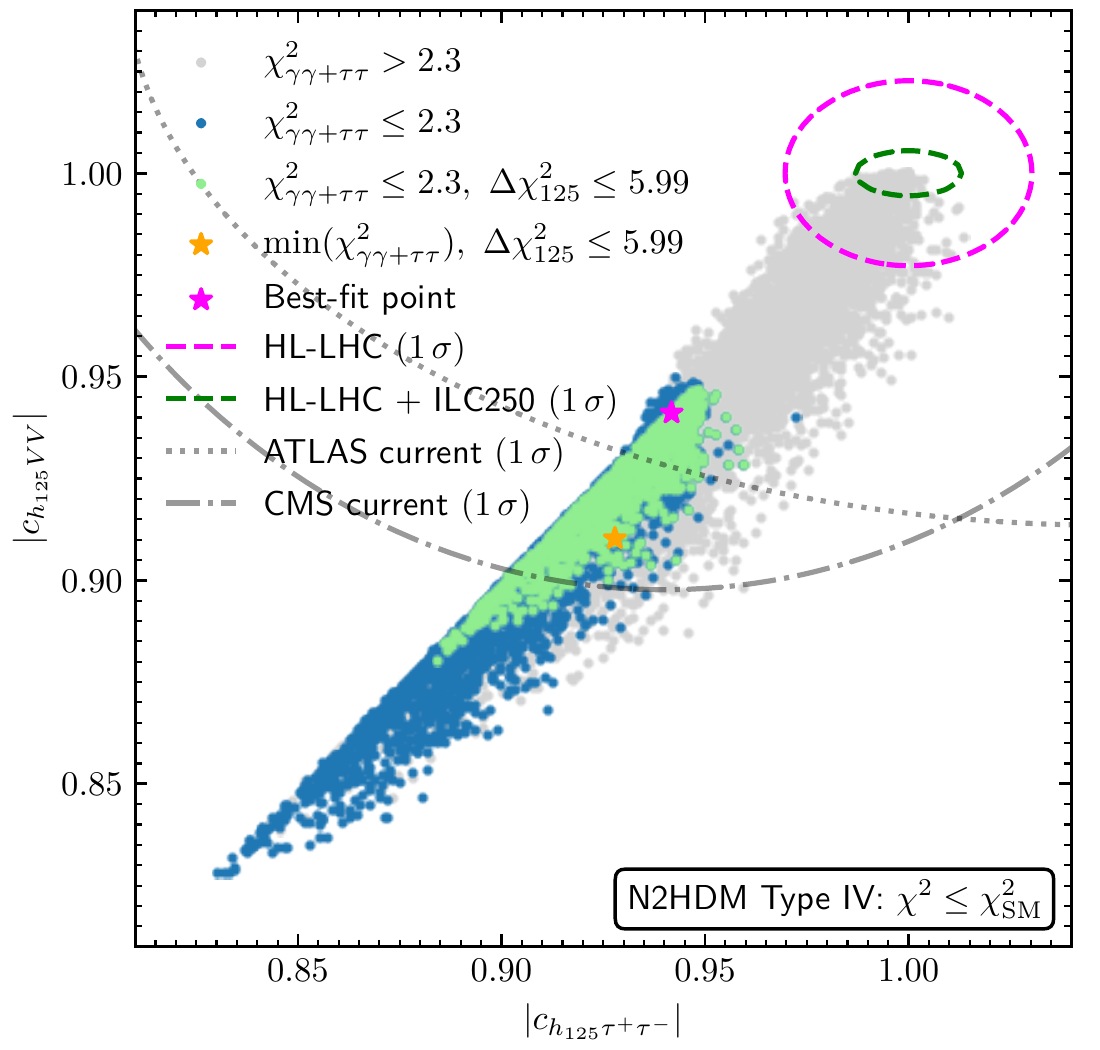}~
\includegraphics[width=0.48\textwidth]{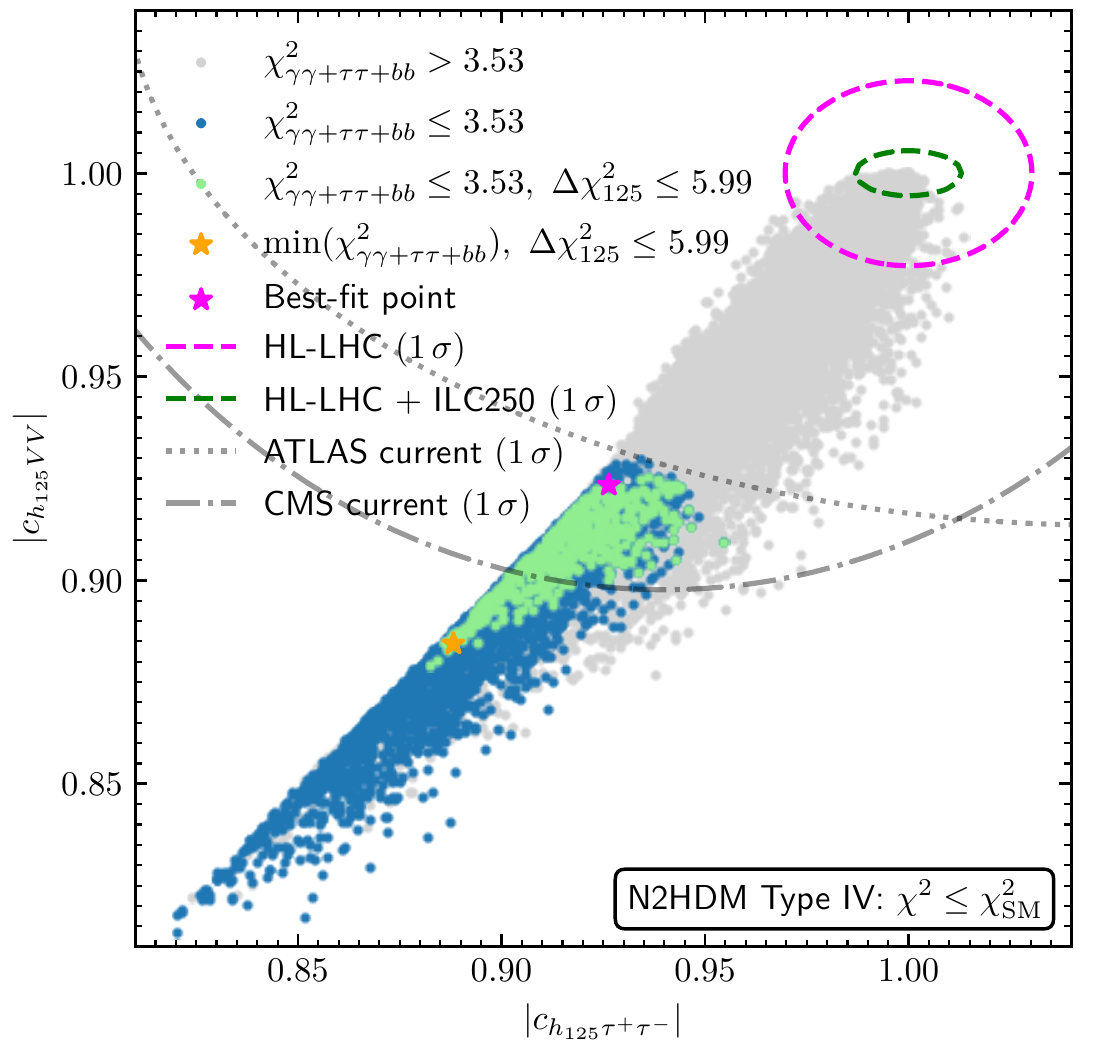}
\caption{\small The $|c_{h_{125}VV}|$--$|c_{h_{125}\tau\tau}|$ 
plane for the parameter
points discussed in \refse{sec:numana1} in
the left \GWn{plot and for the parameter
points discussed} in \refse{sec:numana2} in the
right plot, respectively.
The grey points feature values of
$\chi^2_{\gamgam+\tau\tau} > 2.3$ (left)
and $\chi^2_{\gamgam+\tau\tau+bb} > 3.53$
(right), whereas the blue points feature
$\chi^2_{\gamgam+\tau\tau} \leq 2.3$ \GWn{(left)}
and $\chi^2_{\gamgam+\tau\tau+bb} \leq 3.53$ \GWn{(right)}.
The green points are a subset of the
blue points that \GWn{furthermore} feature
$\Delta \chi^2_{125} \leq 5.99$.
The magenta and orange stars are defined as
in \reffi{fig:muyy_mull_mubb}.
Also shown are the current $1\,\sigma$
uncertainties of the measurements of the
coupling coefficients from ATLAS~\cite{ATLAS:2019nkf}
and CMS~\cite{CMS-PAS-HIG-19-005} indicated by the
dotted and the dash-dotted lines,
respectively.
\GWn{The} magenta and the green ellipse
indicate the prospects for these uncertainties
after the high-luminosity phase of the
LHC~\cite{Cepeda:2019klc}
and after a hypothetical ILC run at a center-of-mass
energy of $250\gev$,
respectively~\cite{Bambade:2019fyw}.}
\label{fig:cplprosp}
\end{figure}

In \reffi{fig:cplprosp} we show
the effective coupling coefficients
$c_{h_{125}\tau\tau}$ and $c_{h_{125}VV}$
for the parameter points discussed in
\refse{sec:numana1} (i.e., the points fitting the 
$\gamma\gamma$ and the $\tau^+\tau^-$ excesses) in the left plot and
for the scan points of the discussion
in \refse{sec:numana2} (i.e., also fitting the $b\bar b$ excess)
in the right plot.
The parameter points are shown in three
different colors. The grey points do not
provide a fit to the excesses that were
considered in each scan within the
$1\sigma$ region of the respective
$\chi^2$-function. On the other hand,
the parameter points depicted in blue
describe the excesses at a level of
$1\sigma$ or below, i.e.~the blue points in the
left plot feature $\chi^2_{\gamgam+\tau\tau}
\leq 2.30$, and the blue points in the
right plot feature $\chi^2_{\gamgam+\tau\tau+bb}
\leq 3.53$.
The points shown in green
(as a subset of the blue points)
furthermore fulfill
the condition $\Delta \chi^2_{125} \leq 5.99$.
The plots also show the current $1\sigma$
uncertainties of the measurements of the
coupling coefficients from ATLAS~\cite{ATLAS:2019nkf}
and CMS~\cite{CMS-PAS-HIG-19-005} indicated by the
dotted and the dash-dotted lines,
respectively. Furthermore, the plots
contain the magenta and the green ellipse
which indicate the prospects for these uncertainties
after the high-luminosity phase of the
LHC~\cite{Cepeda:2019klc}
and after a hypothetical ILC run at a center-of-mass
energy of
$250\gev$~\cite{Bambade:2019fyw}, respectively.
The ellipses are placed such that their
center lies at the SM prediction in order
to visualize the deviations of the couplings
predicted by the scan points compared to
the SM prediction. However, we emphasize that
the placement of the ellipses
is based on a hypothetical scenario
in which the future experiments
will measure no deviations from the SM.

One can see that in both plots the blue
points lie a 
\GWn{significant}
amount away from the
magenta ellipse. As a consequence, independently
of whether the LEP excess is considered or not,
the scenarios that describe the CMS excesses
predict modifications of the
couplings of $h_{125}$ compared to the SM
prediction that would be observable at the
HL-LHC. The discrepancy to the SM predictions
would be even more striking at the ILC. 
\GWn{Regarding} a future lepton-collider it should
\GWn{also} be taken into account that there the state
at $95\gev$ could be 
\tho{probed} directly, such that
the indirect constraints from the properties
of $h_{125}$ 
\GWn{would complement the results from the direct 
search for the state at $95\gev$. In fact,} 
the interplay
between the \GWn{results for the} 
couplings of $h_{95}$ and $h_{125}$
\GWn{will be essential in order to determine}
which
underlying model could be realized in nature.
In order to achieve this goal, the higher precision
of the coupling measurements of $h_{125}$
\tho{at a future $e^+ e^-$ collider}
\GWn{will be} crucial.

We now turn to the direct searches for
BSM Higgs bosons in \GWn{the scenarios that we consider in this paper}.
The fact that one can continue
to search for the state at $95\gev$
in the channels in which
the excesses were observed is
self-evident. However, there is also
the interesting possibility to
shed light on the presence of $h_{95}$
in a complementary way via
searches for the heavier states
$h_3$, $A$ and $H^\pm$.
We found that it is most promising to search
for the third CP-even state $h_3$ due to the
preferred \GWn{relatively} 
low values of its mass (see the
discussion in \refse{sec:numana1}).
The searches for neutral Higgs bosons decaying
into a pair of top quarks are sensitive to the presence
of $h_3$ for values of $\tan\beta \approx 1$ and
masses of $h_3$ not too far above the $t \bar t$
threshold~\cite{CMS:2019pzc}.\footnote{A local excess of
$3.5\,\sigma$ has been observed
in \citere{CMS:2019pzc} for masses of about $400\gev$, which
coincides with the preferred mass range of
$h_3$ found in this analysis. However,
the shape of the excess favours an interpretation
in the form of a \GWn{CP-odd Higgs boson}
decaying into a top-quark
pair (see \citere{Biekotter:2021qbc} for an N2HDM interpretation).}
\GWn{Searches for the} state $h_3$ with \GWn{a mass}
in the range $400\gev \lesssim m_{h_3}
\lesssim 700\gev$ can also be \GWn{carried out}
in the gluon-fusion production mode
with subsequent decay
into a pair of vector bosons
in the low-$\tan\beta$ range~\cite{ATLAS:2018sbw,
CMS:2018amk}.
In addition to these conventional
collider signatures, there is also the possibility
for collider signatures that involve two
BSM states. The combined constraints from flavour-physics
observables and the EWPO give rise to a mass hierarchy
of the form $m_{h_3} < m_{A} \approx m_{H^\pm} \approx 650\gev$.
This hierarchy allows for Higgs cascade decays of the form
$A \to Z h_3$ and $H^\pm \to W^\pm h_3$,
whose branching ratios can be
sizable even in the alignment limit of
the N2HDM. The CP-odd state $A$ can be produced
for small values of $\tan\beta$ in
the gluon-fusion production mode
and for large values of $\tan\beta$ in
the $\bbbar$-associated
production mode. As a result,
future LHC searches utilizing
the signature $A \to h_3 Z$ could probe the N2HDM
type~IV scenario over the whole scan range
of $\tan\beta$. Incidentally, the mass hierarchy
$m_{h_3} < m_{A} \approx m_{H^\pm}$
in combination with low values
of $\tan\beta$ can also facilitate
the realization of \GWn{a} first-order electroweak phase
transition
and electroweak baryogenesis
in the N2HDM~\cite{Biekotter:2021ysx}.\footnote{A
first-order phase
transition can also give rise to a 
gravitational-wave background that might be detectable
in the future.}
Finally, for the lower scan range of $m_{h_3}$
also the decay mode $h_3 \to h_{125} h_{125}$
becomes important. This
signature has been searched for in the final state
with four $b$-quarks~\cite{ATLAS:2018rnh,ATLAS:2019qdc}
and in a final state with
a pair of $b$-quarks and a diphoton
pair~\cite{ATLAS:2018dpp}, and can be further
probed at the (HL-)LHC.

\section{Conclusions}
\label{sec:conclu}

We analyzed local excesses of
\GW{about} $3\,\sigma$ each in the di-photon and
the di-tau decay modes \GW{near} $95\gev$
as reported by CMS, \GW{by themselves 
and} together with
a \GW{long-standing} $2\,\sigma$ local excess in the
$b \bar b$ final state \GW{that was observed}
at LEP in a mass range \GW{that turns out to be}
compatible with the excesses observed by CMS.\footnote{The
fact that all three excesses are found roughly
at the same mass value should reduce the impact
of the ``look-elsewhere'' effect (the difference
between local and global significance).}
\GW{We have investigated whether the observed excesses
could be interpreted as arising from}
a Higgs boson in the
2~Higgs Doublet Model with an additional
real Higgs singlet (N2HDM).
\GW{While in a previous analysis~\cite{Biekotter:2019kde}
it had been found
that the N2HDM of 
type~II and type~IV can describe the
$\gamgam$- and the $\bbbar$-excess
simultaneously, 
we have found that the incorporation of the 
recently
observed excess in the $\tautau$ decay mode
can be well accommodated in the N2HDM but yields a 
clear preference for the type~IV Yukawa structure.}

\begin{figure}
\centering
\includegraphics[width=0.33\textwidth]{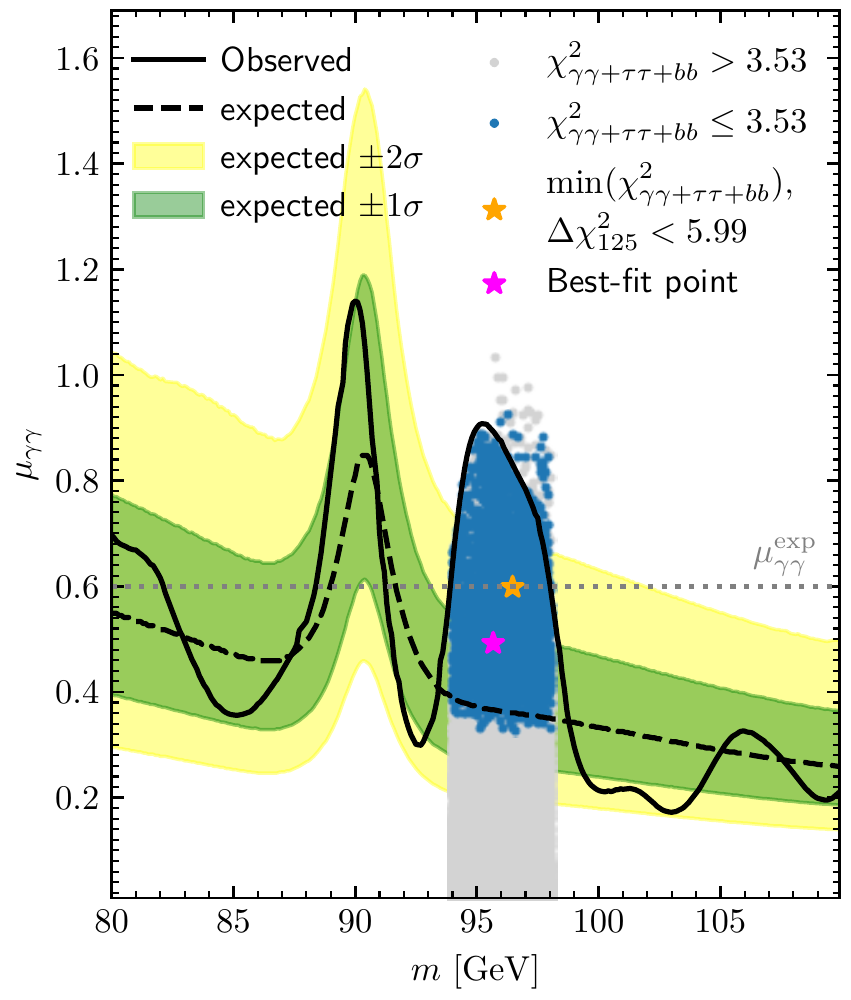}~
\includegraphics[width=0.33\textwidth]{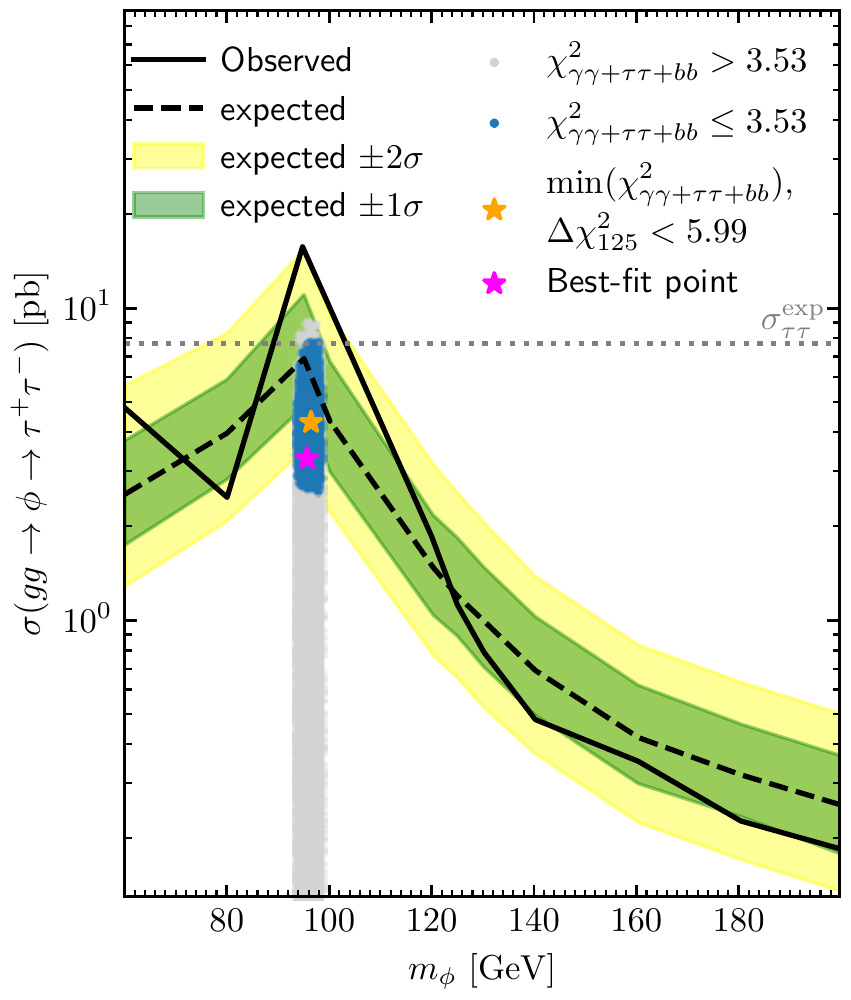}~
\includegraphics[width=0.33\textwidth]{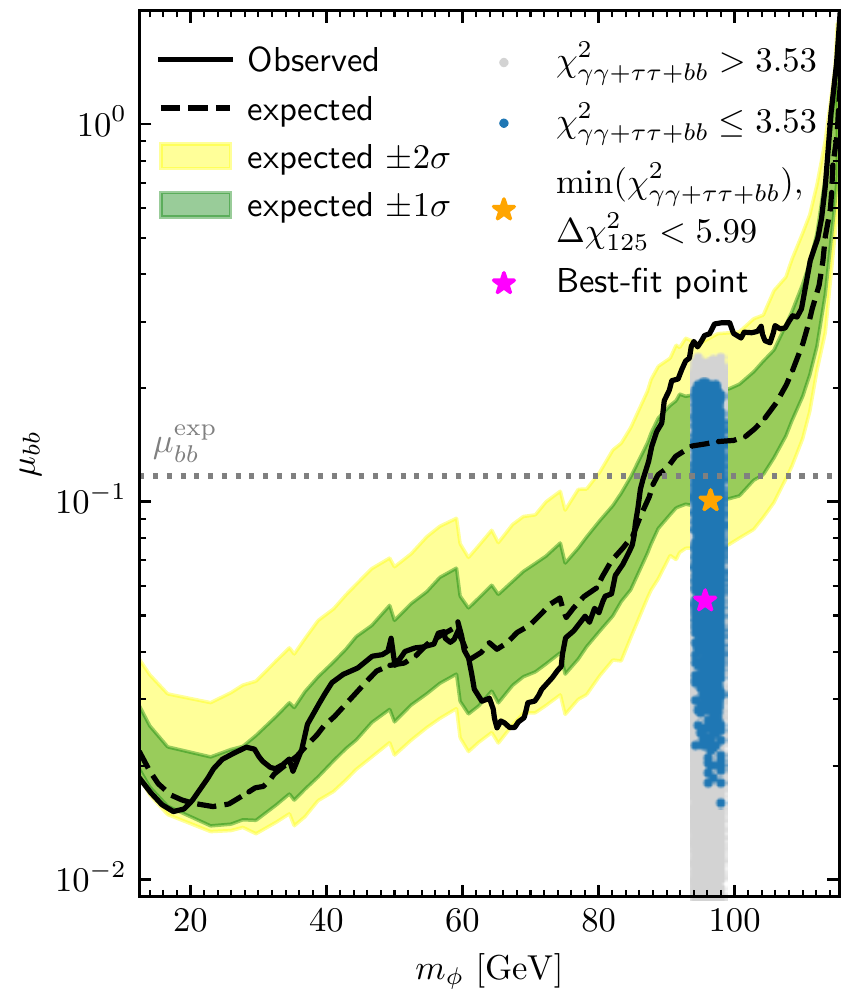}
\caption{\small
Predicted signal rates $\mu_{\gamgam}$
and $\mu_{bb}$
(left and right plots) and
cross sections $\sigma_{\tau\tau}$
(middle plot) with regards to the
three observed excesses.
Blue points feature $\chi^2_{\gamgam+\tau\tau+bb} \leq 3.53$
and describe the excesses at the
level of $1\,\sigma$ or better,
whereas the grey points feature
$\chi^2_{\gamgam+\tau\tau+bb} > 3.53$.
The magenta and orange stars are defined as
in \reffi{fig:muyy_mull_mubb}.
Also shown are the experimentally
observed and expected upper limits at
the 95\% confidence level.
The green and yellow bands indicate the 68\%
and 95\% intervals for
the expected exclusion limit.
The horizontal grey dotted lines indicate
the central values of the observed signal
strengths/cross section of the
three excesses.}
\label{fig:sum}
\end{figure}

\GW{Regarding a simultaneous description of the
$\gamgam$-excess and the $\tautau$-excess
by means of a singlet-like Higgs}
boson with a mass of about $95\gev$,
we have found that the type~IV
Yukawa structure allows a very good description 
\GW{while being in 
agreement with the measurements of the properties of 
the observed Higgs state at $125\gev$ and 
further experimental and theoretical constraints.
On the other hand, 
in the N2HDM type~II a simultaneous description of the excesses in the $\gamgam$ and $\tautau$ channels} is
not possible.
Focusing on the type~IV N2HDM, we
have demonstrated in a second step that
one can accommodate also the $\bbbar$-excess
observed at LEP together with
the two excesses observed at CMS.
As a summary of this result, we show
in \reffi{fig:sum} the signal rates of
the state at $95\gev$ for the three
collider processes in which the excesses
were found including the corresponding
expected and observed 95\%
confidence-level exclusion limits
of each search~\cite{Sirunyan:2018aui,
Barate:2003sz,CMS-PAS-HIG-21-001}.
Here the parameter points that describe
the three excesses at a level of $1\,\sigma$
or better are shown in blue, whereas the
remaining parameter points are shown in grey.
One can see that a
large \GW{set} of parameter points
yields a good description of the observed
excesses, while being in agreement with
the theoretical and experimental constraints
on the model parameters.
In particular, we have verified that
a description of the excesses is possible
without \GW{large} modifications of the
measured signal rates of the Higgs boson
at $125\gev$ that would be in disagreement
with the current experimental results.

There are various ways by which future
collider experiments can shed \GW{more} light
on whether the \GW{observed experimental ``anomalies''}
at $95\gev$
\GW{have indeed} a BSM particle origin.
Direct searches for the state at $95\gev$
at the LHC in the $\gamgam$ and the
$\tautau$ final states will
obviously be crucial.
We are eagerly awaiting the updated results
in the $\gamgam$ decay mode by CMS
utilizing the full Run~2 dataset.
If further evidence will be present in this
search after the inclusion of the
remaining Run~2 data, the combined statistical
significance of the various excesses
could indicate 
\GW{a striking indication} for new physics.
With regard to the $\tautau$ decay mode,
we note \GW{once more} 
that ATLAS results in the low-mass Higgs-boson
searches utilizing the $\tautau$
decay mode are yet to be published.
In addition to the direct searches,
we emphasized that \GW{improved measurements of
the properties of the observed Higgs
boson at $125\gev$ have the potential to}
probe the N2HDM scenarios
presented here. The future precision of the
couplings measurements of $h_{125}$
at the high-luminosity phase
of the LHC will be sufficient
to exclude or confirm the N2HDM scenarios
\GW{discussed} here with respect to
the SM at the 95\% confidence-level or more.
At 
\GWn{an $e^+e^-$ Higgs factory} 
\hto{running at $\sqrt{s} = 250 \gev$}
the state
$h_{95}$ could be produced copiously,
and the measurements of the couplings of
both the state at $95\gev$ and $125\gev$
could shed further light on the underlying
model and its parameter space.
Finally, we discussed that there are good
prospects for a discovery of one or more of the
heavier Higgs bosons at the HL-LHC.

\subsection*{Acknowledgements}

The work of S.H.\ is supported in part by
the grant PID2019-110058GB-C21 funded by
``ERDF A way of making Europe'' and by
MCIN/AEI/10.13039/501100011033, and in part
by the grant CEX2020-001007-S funded by
MCIN/AEI/10.13039/501100011033.
The work of T.B.\ and G.W.\ is supported by the Deutsche
Forschungsgemeinschaft under Germany’s Excellence Strategy EXC2121
“Quantum Universe” - 390833306.
\tho{This work has been partially funded by
the Deutsche Forschungsgemeinschaft 
(DFG, German Research Foundation) - 491245950.}

\appendix

\tho{
\section{Additional application of
\hto{the \GWn{limits from the} LEP $\tau^+ \tau^-$} searches}
\label{sec:append}

\GWn{As discussed in the context of our results shown
in \reffi{fig:muyy_mull_chisqhs_4} and 
\reffi{fig:muyy_mull_mubb}, 
parts of the parameter space investigated
in our numerical analyses would be excluded
if one additionally had demanded agreement with
the cross-section limits resulting from
the searches for $e^+ e^- \to h_{95} \to
\tau^+ \tau^-$ at LEP~\cite{Barate:2003sz}
for each parameter point, independently
of whether this search was selected as the most
sensitive search based on the expected limits
following the approach implemented
in \texttt{HiggsBounds}.
Applying the
exclusion limits from this experimental search
in combination with the application of the exclusion
limits of the search that was selected by
\texttt{HiggsBounds} effectively yields
a limit that is stronger than 
an overall 95\% C.L..}
\GWn{In our analysis above we followed the 
approach to maintain 
a statistical interpretation of
the applied cross-section limits from
BSM Higgs-boson searches
as an overall exclusion bound at the 
95\% C.L.\ and therefore did not impose the 
limits from the LEP searches in the $\tau^+ \tau^-$ final states as additional constraint.
However,}
in order
to demonstrate the potential impact of 
\GWn{those limits} 
on the parameter space favoured by the
excesses at $95\gev$, we show in
\reffi{fig:referee1} \hto{(corresponding to the right plot 
of \reffi{fig:muyy_mull_chisqhs_4})} and \reffi{fig:referee2} \hto{(corresponding to 
\reffi{fig:muyy_mull_mubb})}
the signal rates of the state 
$h_{95}$, 
where the parameter points that predict a
cross section larger than the LEP \hto{$\tau\tau$} limit are shown
in grey. 
\GWn{As explained above,}
the parameter
points shown in \reffi{fig:referee1} were obtained
based on the condition shown in
\refeq{eq:nolep}, such that the LEP excess in the
$h_{95} \to b \bar b$ decay mode is not taken into
account, whereas the parameter points shown
in \reffi{fig:referee2} were obtained based
on the condition shown in \refeq{eq:yeslep}
that includes the contribution $\chi^2_{b b}$.

\begin{figure}
\centering
\includegraphics[width=0.48\textwidth]{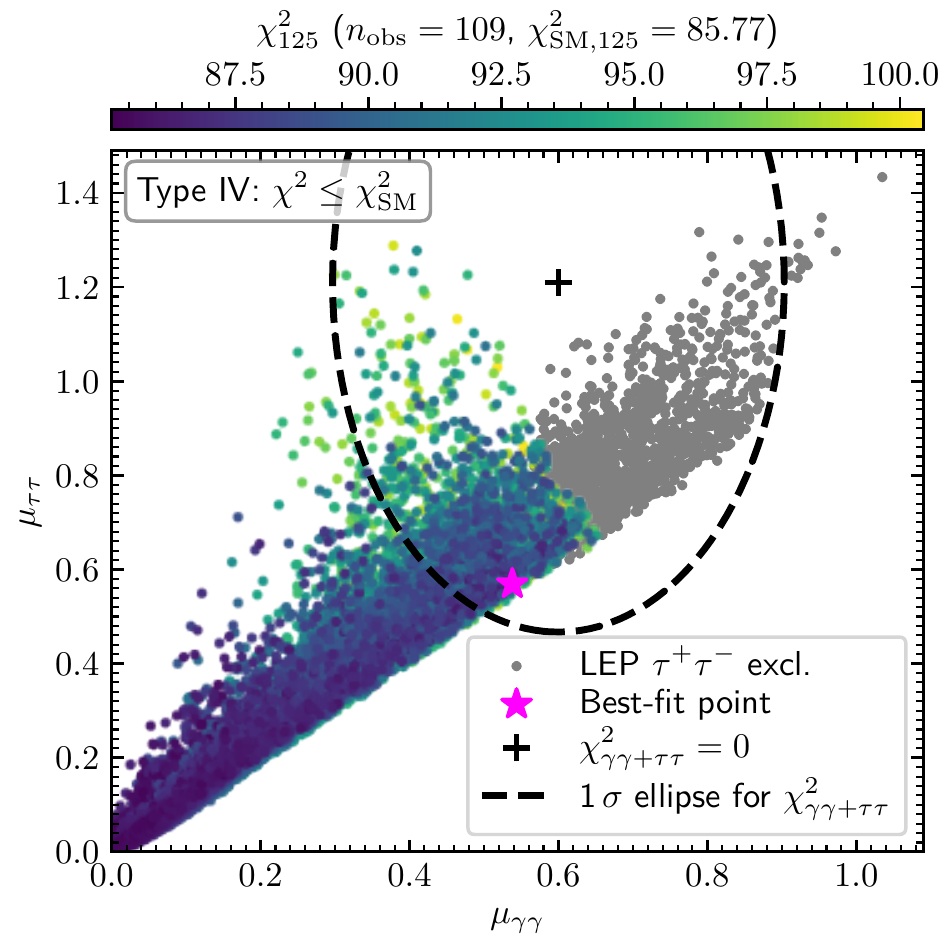}
\caption{\small As in the right plot
of \reffi{fig:muyy_mull_chisqhs_4}, but
\GWn{those} parameter points 
are shown in grey (plotted below the
colored points)
\GWn{that would be excluded 
by the LEP limit in the $\tau^+\tau^-$ final state 
if that channel had been selected in order to 
determine the 95\% C.L.\ limit.}
}
\label{fig:referee1}
\end{figure}

\begin{figure}
\centering
\includegraphics[width=0.48\textwidth]{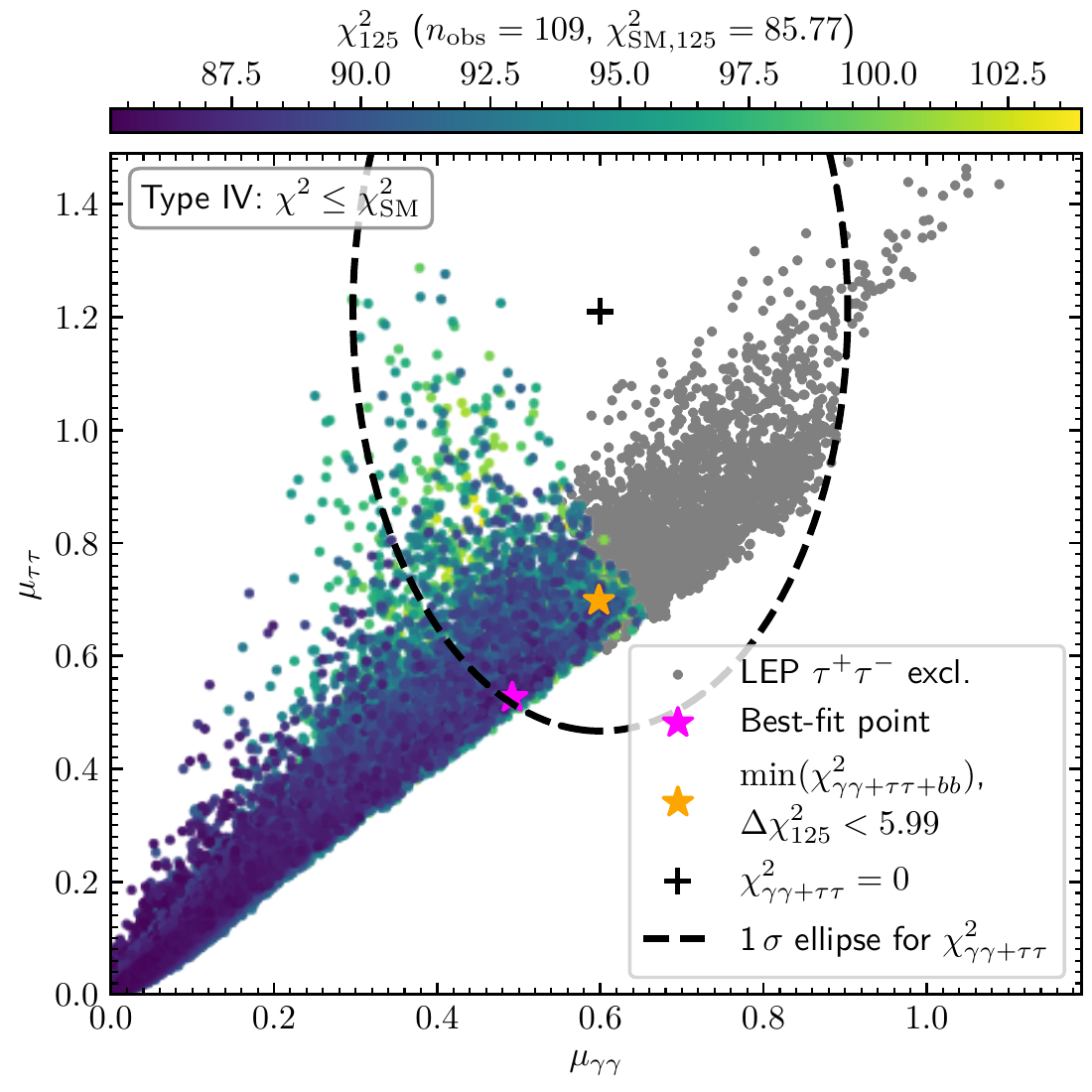}
\includegraphics[width=0.48\textwidth]{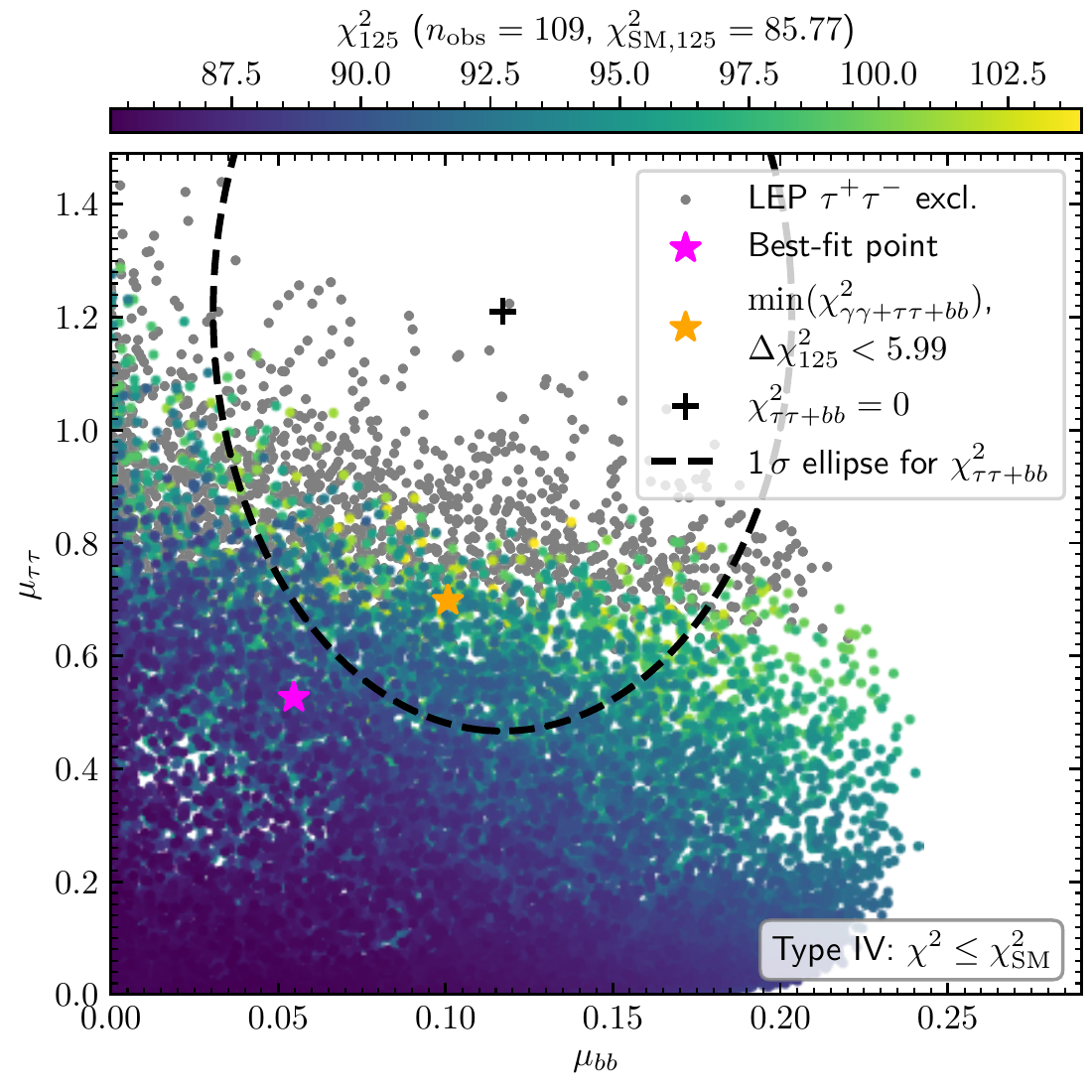}
\caption{\small As in
\reffi{fig:muyy_mull_mubb}, but
\GWn{those} parameter points are shown in grey 
(plotted below the
colored points)
\GWn{that would be excluded 
by the LEP limit in the $\tau^+\tau^-$ final state 
if that channel had been selected in order to 
determine the 95\% C.L.\ limit.}
}
\label{fig:referee2}
\end{figure}

}

\bibliography{lit}

\end{document}